\documentclass{emulateapj}

\usepackage{psfig}
\usepackage{amsmath}
\usepackage{amssymb}
\usepackage{graphicx}
\usepackage{appendix}
\usepackage{color}

\newcommand{\lSect}[1]{{\label{sec:#1}}}
\newcommand{\lFig}[1]{{\label{fig:#1}}}

\newcommand{\lTab}[1]{{\label{tab:#1}}}
\def\gtaprx {\lower .1ex\hbox{\rlap{\raise .6ex\hbox{\hskip .3ex
	{\ifmmode{\scriptscriptstyle >}\else
		{$\scriptscriptstyle >$}\fi}}}
	\kern -.4ex{\ifmmode{\scriptscriptstyle \sim}\else
		{$\scriptscriptstyle\sim$}\fi}}}
\def\ltaprx {\lower .1ex\hbox{\rlap{\raise .6ex\hbox{\hskip .3ex
	{\ifmmode{\scriptscriptstyle <}\else
		{$\scriptscriptstyle <$}\fi}}}
	\kern -.4ex{\ifmmode{\scriptscriptstyle \sim}\else
		{$\scriptscriptstyle\sim$}\fi}}}
\newcommand{\FIGFF}[2]{{\ref{fig:#2}{#1}}}

\newcommand{\FIG}[2]{{Fig.~\FIGFF{#1}{#2}}}
\newcommand{\Fig}[1]{{\FIG{}{#1}}}

\newcommand{\Sectff}[1]{{\ref{sec:#1}}}
\newcommand{\Sect}[1]{{\S~\Sectff{#1}}}

\newcommand{\Msun}{\ensuremath{\mathrm{M}_\odot}}
\newcommand{\Lsun}{\ensuremath{\mathrm{L}_\odot}}

\newcommand{\Zsun}{\ensuremath{\mathrm{Z}_\odot}}
\newcommand{\Tab}[1]{{Table \ref{tab:#1}}}

\usepackage{soul}

\begin{document}

\title{SN 1961V: A Pulsational Pair-Instability Supernova}

\author{S. E. Woosley\altaffilmark{1} and Nathan Smith\altaffilmark{2}}

\altaffiltext{1}{Department of Astronomy and Astrophysics, University
  of California, Santa Cruz, CA 95064; woosley@ucolick.org}
\altaffiltext{2}{Steward Observatory, University of Arizona, 933 N. Cherry
  Ave., Tucson, AZ 85721, USA}

\begin{abstract} 
We explore a variety of models in which SN~1961V, one of the most
enigmatic supernovae (SNe) ever observed, was a pulsational
pair-instability supernova (PPISN).  Successful models reproduce the
bolometric light curve of the principal outburst and, in some cases,
the emission one year before and several years afterward.  All models
have helium-rich ejecta, bulk hydrogenic velocities near 2000 km
s$^{-1}$, and total kinetic energies from 4 to 8 $\times 10^{50}$
erg. Each eventually leaves behind a black hole remnant. Three
subclasses of PPISN models are explored, each with two different
choices of carbon abundance following helium burning. Carbon is an
important parameter because shell carbon burning can weaken the
explosion. The three subclasses correspond to situations where
SN~1961V and its immediate afterglow were: a) a single event; b) the
first of two or more pulsational events separated by decades or
centuries; or c) the latter stages of a complex explosion that had
already been going on for a year or more.  For the low carbon case,
the main sequence mass for SN~1961V's progenitor would have been 100
to 115 \Msun; its pre-SN helium core mass was 45 to 52 \Msun; and the
final black hole mass, 40 to 45 \Msun. For the high-carbon case, these
values are increased by roughly 20 to 25\%. In some PPISN models, a
$\sim10^{40}$ erg s$^{-1}$ star-like object could still be shining at
the site of SN~1961V, but it has more likely been replaced by a
massive accreting black hole.
\end{abstract}


\section{INTRODUCTION}
\lSect{intro}

Few supernovae (SNe) have histories rivaling the rich, controversial
case of SN 1961V in the nearby ($\sim$10 Mpc) spiral galaxy NGC~1058.
Discovered in July, 1961 by \citet{Wil61}, the event was originally
classified by \citet{Zwi64} as a Type V SN. It was, for a time,
the prototypical ``supernova impostor'', noted, in particular, for its
similarity to Eta Carinae \citep{Zwi64,Goo89,Fil95,Hum99}. In fact,
Eta Carinae was Zwicky's only other Type V SN. The progenitor was
first seen as a 19th magnitude object starting in 1937, and repeatedly
visited by observers thereafter \citep{Ber64,Zwi64}. An appreciable
rise in brightness by roughly a factor of 10 was reported in
November 1960, about a year before the SN.  During the ``main event''
in late 1961, the irregular luminosity exceeded, for a time, that of a
typical Type~II SN.  Overall, the light curve of this main event was
extremely unusual among known SNe, with a $\sim$100 day plateau
resembling a typical SN II-P at an absolute magnitude of roughly
$-$16.5 mag, punctuated by a late, brief luminosity spike at almost
$-$18 mag.  The decline from this last peak was slow and unsteady over
the subsequent decade.

Given the large inferred luminosity of the star before it exploded,
\citet{Goo89} suggested a zero-age main-sequence (ZAMS) mass of over
240 \Msun \ and a pre-outburst mass of over 170 \Msun.  Their
estimated absolute magnitude, $-$12.1 mag, relied on an uncertain
extinction correction of 0.6 mag though, and the star could have been
fainter, or multiple.\footnote{Although this is also distance
  dependent, and modern studies favor a somewhat larger distance than
  adopted by Goodrich et al.} \citet{Koc11} later reduced the lower
limit for the mass to 80 \Msun \ and assigned a metallicity $\sim$1/3
solar.  Even then, the SN progenitor was one of the brightest stars in
NGC 1058 and clearly quite massive. A very high-mass progenitor is
reasonable based on its local environment, since SN~1961V resides
within a young giant H~{\sc ii} region
\citep{Goo89,Fil95,Van02,Chu04}. The measured helium to hydrogen ratio
in the SN ejecta was at least 4 times solar
\citep{Bra71}. Observations showed that the SN returned to
approximately its pre-SN brightness from about 1.5 to 4 years after
its peak \citep{Ber70}, but then continued to fade steadily
afterward. By seven years after its peak, the source at SN~1961V's
position declined to $\sim$3 mag fainter than the progenitor at
visual wavelengths \citep{Ber70}, and after 4 decades it had faded to
about 7 mag fainter than the progenitor.

The SN, if it was one, definitely had some unexpected physical
characteristics \citep{Ber64,Zwi64,Bra71}. Its expansion velocity,
measured from line widths in its spectra, was unusually slow compared
to most SNe II of similar luminosity. \citet{Zwi64} reported a speed,
based on H$\alpha$ near peak, of about 3700 km s$^{-1}$, but noted the
presence of many narrower lines and commented that the ``true
expansion speed'' was probably slower. \citet{Bra71} estimated
$\sim2000$ km s$^{-1}$ from fitting the entire spectrum, and
\citet{Goo89} estimated 2100 km s$^{-1}$ from analysis of H$\alpha$ at
late times. The spectrum was not like that of common Type I or II SNe
\citep{Bra71}. If the same event were discovered today, it would most
likely be classified as Type~IIn, but that designation did not exist
at the time (see \citealt{Smi11c}). The main light curve shape was
unusual and irregular. The SN had long-lasting precursor emission, and
its post-peak decline lasted many years, with several late peaks and
plateaus as it faded.  The inferred energetics were also unusual. For
an ejecta mass around 10 \Msun, the low velocity implied a low kinetic
energy, $\sim 4 \times 10^{50}$ erg.  This is less than the
energy of a normal core-collapse SN, but somewhat higher than the
inferred energy budget of Eta Carinae's Great Eruption \citep{Smi03}.
While it had relatively low kinetic energy and was faint except for
its brief maximum, the SN's duration implied a large amount of total
radiation.  \citet{Bra71} estimated that the 1961 outburst had an
integrated radiated energy of $\sim 2 \times 10^{50}$ erg. That is,
even with substantial errors for distance and bolometric correction,
it gave off more light than that radiated by a typical core-collapse
SN. Such a large light to kinetic energy ratio is suggestive of a more
efficient conversion than in common SN. This, in turn, suggests that
the luminosity of SN~1961V was powered, in part, by shock interaction
with dense circumstellar material (CSM) or multiple shell collisions
\citep{Smi11c}.

The assumption that SN 1961V was a core-collapse SN has frequently
been questioned. Often arguing by analogy to Eta Carinae, the
outbursts and irregular behavior have been attributed to some sort of
stellar variability, often lumped under the rubric of ``the outburst
of a luminous blue variable (LBV)'' \citep{Goo89,Hum99}. A
decades-long debate has raged as to whether the star is still there
\citep{Van02,Fil95,Chu04,Smi11c,Koc11,Van12}. A common assumption,
which shall later be questioned (\Sect{long}), is that, if the star is
still there, a SN did not occur.  A key point is that if the
progenitor was not completely obliterated by the 1961 event, then this
rules out a traditional Fe core-collapse SN, but it does not rule out
other SN explosion mechanisms like the one we consider here. The LBV
hypothesis has in its favor the long and varied history of SN~1961V
and its progenitor, which included some outbursts that were well below
typical SN luminosities; the low velocity; the peculiar Eta Car-like
spectrum; and the temporary return of the luminosity after the 1961
outburst, to values similar to the pre-SN star.  On the other hand,
the mechanism that drives giant LBV eruptions is not known, and
SN~1961V was far more extreme in terms of luminosity and expansion
velocity than other LBV eruptions \citep{Smi11c}.

More recently, it has been looking less likely that the
pre-SN star survived, though the subject certainly remains
controversial and relevant to any modeling. A fading radio source
typical of a SN was discovered at the site of SN 1961V,
suggesting that this was not just the eruption of a variable star
\citep[e.g.][]{Bra85,Chu04}. \citet{Pat19} placed limits on dust
obscuration and concluded that, at the present time, any survivor must
be much less luminous than the pre-SN star prior to its
outburst.  \citet{Smi11c} have argued that the velocity and energetics
of SN 1961V are not unusual for a SN of Type IIn. See also
\citet{Koc12}.

Here we adopt the working hypothesis that SN~1961V was a Type~IIn SN,
i.e. the explosive death of a massive star with a residual hydrogen
envelope and dense CSM. That death may have been novel and might have
taken a long time, but the star is either dead (\Sect{single} and
\Sect{multiple}) or terminally ill (\Sect{long}). A key question then
is, if it was a SN, how did it explode?  A traditional Fe core
collapse with a single neutrino-driven explosion is not the only way
that a massive star can die.  Neutrino transport is thought to be
insufficiently effective to explode such massive stars
\citep{Rah22}. A previous SN-based model sidestepped this issue and
suggested that SN~1961V was the explosion of a supermassive 2000 \Msun
\ star \citep{Utr84}. Another suggested that SN~1961V was a more
common SN (presumably of lower mass) experiencing an unusual amount of
CSM interaction \citep{Smi11c}. Both attempted to explain the long
duration of the event and its complex time history, the former by a
layered structure for the exploding star \citep{Utr84}, the later by a
variable circumstellar density \citep{Smi10,Smi11c}. Aside from the
obvious issue of whether 2000 \Msun \ pre-SN stars exist, there is no
known mechanism for exploding them. They are expected to collapse
directly to black holes \citep{Heg02,Heg03,Rah22}.  With rotation and
progenitors derived from a stellar evolution code, neither the layered
structure nor the explosion of a 2000 \Msun \ model resembled the
parametric models of Utrobin (see \citet{Ens89} and Stringfellow as
cited in \citet{Goo89}). The model of \citet{Smi11c} was qualitative
and lacked a specific explosion or stellar model, but resembles in
outcome, what will be discussed here.

Here we explore the consequences of models in which SN 1961V exploded
as a pulsational pair-instability supernova
\citep[PPISN;e.g.][]{Bar67,Heg02,Woo07,Woo17}. \citet{Pas13} have also
speculated that SN 1961V may have been a PPISN, but no specific
model was presented. This mechanism can produce long duration,
irregular light curves in stars having $M_{\rm ZAMS}$ comparable to
the inferred pre-SN mass. The luminosity of the pre-SN
star is roughly consistent with 23 years of observations and, in some
cases, the explosion returns to approximately its pre-SN value
after the major outburst. The velocity of most of the hydrogen-rich
ejecta of PPISN is near 2000 km s$^{-1}$ for reasonable hydrogenic
envelope masses.  PPISN are expected to occur in regions with low
metallicity, mainly because of the prodigious mass-loss winds of such
luminous stars.  The restrictions on low metallicity might be relaxed
somewhat, however, due to modern reductions in empirical mass-loss
rates \citep{Smi14}.

Unlike traditional core-collapse SNe --- but similar to SNe IIn --
shock interaction with matter ejected in previous pulses or by the
pre-SN wind can, in some cases, contribute appreciably to the light
curve.  Many solar masses of material are available for CSM
interaction, and shells typically interact between 10$^{15}$ cm and
10$^{16}$ cm. The helium to hydrogen ratio is naturally very
supersolar, and the efficiency for converting kinetic energy to light
is high. These are events that should occur in nature and produce
black holes with masses like those detected by LIGO
\citep[e.g.,][]{Abb19}. Their explosion characteristics are robust and
do not involve neutrino transport.

Complicating our study, though, it turns out that there are three PPISN
solutions to SN 1961V, each with different strengths and
weaknesses. There are models in which: a) The SN was a singular event
in late 1961 lasting of order 200 days during which multiple pulses
occurred and the remnant collapsed to a black hole (\Sect{single}). In
this case, all transient emission before and after the 1961 event
requires other explanations, interaction with a prior wind or LBV-like
eruptions being possibilities as inferred in many SNe IIn, including
the remarkable case of SN~2009ip \citep{Smi10b}.  b) The pulsations
continued for decades or centuries with a long quiescent interval
between. The main outburst of SN 1961V was then just the first one of
two major clusters of pulsations. The second, and an eventual collapse
to a black hole, would happen much later. In this case, something
resembling the pre-SN star could even still be present, along
with a fading radio source (\Sect{long}). c) The SN actually commenced
well in advance of November 1961, and some major bright prior
emission was missed by observers at the time. The enduring emission
after a year (i.e., after 1962) is due to shells continuing to collide
after the SN core has long since collapsed to a black hole
(\Sect{multiple}).  In all cases, a black hole remnant with 40 to 50
\Msun \ is the ultimate product.

\section{The Observed Light Curve}
\lSect{litecurve}

The SN~1961V light curve can be divided into four parts: (1) variable
pre-SN emission prior to July 1961, (2) a plateau similar to
that of a typical SN~II-P, lasting about 120 days, (3) a delayed peak
about four times brighter lasting about 30 days, and (4) a rapid
decline to two additional long plateaus and slow fading thereafter.
The distance to NGC 1058 has historically been uncertain, leading to
large ranges in the estimated SN luminosity in the literature. Based
upon a compromise between similar values of distance moduli deduced
from the expanding photosphere method for SN 1969L, which also
happened in NGC 1058 (m-M = 30.13 to 30.25 mag), and Hubble expansion
(m-M = 29.77), \citet{Smi11c} recommended a distance modulus of 30.0
which will be adopted here, with an uncertainty of about 0.2 mag. An
additional correction for Galactic extinction of 0.3 mag is assumed.

\Tab{61vlite} then gives the apparent photographic magnitudes
\citet{Smi11c} adapted from \citet{Zwi64} and \citet{Ber64} and the
equivalent luminosities, assuming no bolometric correction. Two of the
points, on days 1960.641 and 1960.704 are upper bounds to the
brightness.  The time when the light curve reaches its peak is
1961.943, with a photographic magnitude (approximately equal to the
$B$-band magnitude) of 12.50 mag, corresponding to a peak absolute
magnitude of $-$17.8. All magnitudes prior to 1964.186 have an
observational uncertainty of about 0.1 mag, and points after are
uncertain to 0.2 mag. The peak luminosity, again ignoring bolometric
correction, is
\begin{equation}
  {\rm log}_{10} (L/\Lsun) \ = 0.4 (4.74 - (m_{pg} - 30.3)).
\end{equation}
The luminosity on the plateau is thus near 10$^{42}$ erg s$^{-1}$, and
the peak luminosity is near $4 \times 10^{42}$ erg s$^{-1}$. The
former is typical for SNe~II-P on their plateaus
\citep{Suk16,Smi11c}. Given uncertainties in distance, bolometric
correction and extinction, these estimated luminosities could easily
be off by 50\%.

\begin{deluxetable*}{cccccccccccc} 
\tablecaption{SN 1961V Light Curve} 
\tablehead{Date & to L$_{\rm max}$ & m$_{\rm pg}$ & L & Date & to L$_{\rm max}$ & m$_{\rm pg}$ & L & Date & to L$_{\rm max}$ & m$_{\rm pg}$ & L \\
(year) &  (days) & (mag)  & (erg s$^{-1}$) & (year) & (days) & (mag)  & (erg s$^{-1}$) & (year) &  (days) & (mag)  & (erg s$^{-1}$) } 
\startdata
1937.786 & -8823.3 & 18.20 & 2.10E+40 & 1961.866 &   -28.1 & 13.75 & 1.26E+42 & 1962.805 &   314.8 & 17.50 & 4.00E+40 \\
1937.849 & -8800.3 & 18.20 & 2.10E+40 & 1961.868 &   -27.4 & 13.66 & 1.37E+42 & 1962.806 &   315.2 & 17.60 & 3.65E+40 \\
1946.918 & -5487.9 & 18.00 & 2.52E+40 & 1961.871 &   -26.3 & 13.60 & 1.45E+42 & 1962.816 &   318.9 & 17.55 & 3.82E+40 \\
1946.920 & -5487.2 & 18.00 & 2.52E+40 & 1961.874 &   -25.2 & 13.40 & 1.75E+42 & 1962.817 &   319.2 & 17.40 & 4.39E+40 \\
1949.810 & -4431.6 & 18.10 & 2.30E+40 & 1961.901 &   -15.3 & 13.00 & 2.52E+42 & 1962.820 &   320.3 & 17.70 & 3.33E+40 \\
1949.915 & -4393.2 & 17.90 & 2.77E+40 & 1961.920 &    -8.4 & 12.80 & 3.03E+42 & 1962.827 &   322.9 & 17.50 & 4.00E+40 \\
1951.812 & -3700.3 & 17.70 & 3.33E+40 & 1961.923 &    -7.3 & 13.00 & 2.52E+42 & 1962.838 &   326.9 & 17.56 & 3.78E+40 \\
1951.920 & -3660.9 & 17.70 & 3.33E+40 & 1961.932 &    -4.0 & 12.66 & 3.45E+42 & 1962.839 &   327.3 & 17.54 & 3.85E+40 \\
1952.649 & -3394.6 & 17.70 & 3.33E+40 & 1961.935 &    -2.9 & 12.97 & 2.59E+42 & 1962.874 &   340.0 & 17.65 & 3.48E+40 \\
1954.973 & -2545.8 & 18.00 & 2.52E+40 & 1961.937 &    -2.2 & 12.80 & 3.03E+42 & 1962.890 &   345.9 & 17.64 & 3.52E+40 \\
1960.641 &  -475.6 & $>$17.0 & $<$6.3E+40 & 1961.940 &    -1.1 & 12.70 & 3.33E+42 & 1962.896 &   348.1 & 17.80 & 3.03E+40 \\
1960.704 &  -452.5 & $>$17.0 & $<$6.3E+40 & 1961.943 &     0.0 & 12.50 & 4.00E+42 & 1962.915 &   355.0 & 17.80 & 3.03E+40 \\
1960.942 &  -365.6 & 15.80 & 1.91E+41 & 1961.945 &     0.7 & 12.85 & 2.90E+42 & 1962.918 &   356.1 & 17.73 & 3.24E+40 \\
1961.539 &  -147.6 & 13.50 & 1.59E+42 & 1961.951 &     2.9 & 12.71 & 3.30E+42 & 1962.956 &   370.0 & 18.00 & 2.52E+40 \\
1961.562 &  -139.2 & 13.75 & 1.26E+42 & 1961.954 &     4.0 & 12.90 & 2.77E+42 & 1963.041 &   401.0 & 18.03 & 2.45E+40 \\
1961.589 &  -129.3 & 14.00 & 1.00E+42 & 1961.958 &     5.5 & 13.46 & 1.65E+42 & 1963.066 &   410.2 & 18.10 & 2.30E+40 \\
1961.591 &  -128.6 & 13.90 & 1.10E+42 & 1961.980 &    13.5 & 14.04 & 9.68E+41 & 1963.074 &   413.1 & 18.00 & 2.52E+40 \\
1961.583 &  -131.5 & 13.82 & 1.19E+42 & 1961.986 &    15.7 & 14.20 & 8.36E+41 & 1963.148 &   440.1 & 18.20 & 2.10E+40 \\
1961.594 &  -127.5 & 13.97 & 1.03E+42 & 1962.003 &    21.9 & 14.54 & 6.11E+41 & 1963.153 &   442.0 & 18.34 & 1.85E+40 \\
1961.597 &  -126.4 & 14.06 & 9.51E+41 & 1962.011 &    24.8 & 15.10 & 3.65E+41 & 1963.156 &   443.0 & 18.34 & 1.85E+40 \\
1961.600 &  -125.3 & 13.86 & 1.14E+42 & 1962.016 &    26.7 & 15.25 & 3.18E+41 & 1963.227 &   469.0 & 19.00 & 1.00E+40 \\
1961.602 &  -124.6 & 13.80 & 1.21E+42 & 1962.021 &    28.5 & 15.20 & 3.33E+41 & 1963.551 &   587.3 & 18.30 & 1.91E+40 \\
1961.608 &  -122.4 & 13.92 & 1.08E+42 & 1962.022 &    28.9 & 15.30 & 3.03E+41 & 1963.567 &   593.2 & 18.42 & 1.71E+40 \\
1961.611 &  -121.3 & 13.88 & 1.12E+42 & 1962.033 &    32.9 & 15.35 & 2.90E+41 & 1963.586 &   600.1 & 18.46 & 1.65E+40 \\
1961.625 &  -116.1 & 14.08 & 9.33E+41 & 1962.063 &    43.8 & 15.60 & 2.30E+41 & 1963.636 &   618.4 & 18.41 & 1.73E+40 \\
1961.627 &  -115.4 & 14.02 & 9.86E+41 & 1962.066 &    44.9 & 15.35 & 2.90E+41 & 1963.649 &   623.1 & 18.47 & 1.64E+40 \\
1961.666 &  -101.2 & 13.96 & 1.04E+42 & 1962.068 &    45.7 & 15.60 & 2.30E+41 & 1963.666 &   629.3 & 18.35 & 1.83E+40 \\
1961.672 &   -99.0 & 14.06 & 9.51E+41 & 1962.074 &    47.8 & 15.70 & 2.10E+41 & 1963.764 &   665.1 & 18.47 & 1.64E+40 \\
1961.674 &   -98.3 & 13.98 & 1.02E+42 & 1962.075 &    48.2 & 15.75 & 2.00E+41 & 1963.805 &   680.1 & 18.40 & 1.75E+40 \\
1961.679 &   -96.4 & 13.88 & 1.12E+42 & 1962.079 &    49.7 & 15.70 & 2.10E+41 & 1963.833 &   690.3 & 18.40 & 1.75E+40 \\
1961.681 &   -95.7 & 13.92 & 1.08E+42 & 1962.085 &    51.9 & 15.50 & 2.52E+41 & 1963.868 &   703.1 & 18.40 & 1.75E+40 \\
1961.682 &   -95.3 & 14.00 & 1.00E+42 & 1962.090 &    53.7 & 15.40 & 2.77E+41 & 1963.895 &   713.0 & 18.40 & 1.75E+40 \\
1961.688 &   -93.1 & 14.06 & 9.51E+41 & 1962.095 &    55.5 & 16.00 & 1.59E+41 & 1963.942 &   730.1 & 18.40 & 1.75E+40 \\
1961.690 &   -92.4 & 13.92 & 1.08E+42 & 1962.096 &    55.9 & 16.30 & 1.21E+41 & 1964.011 &   755.3 & 18.45 & 1.67E+40 \\
1961.690 &   -92.4 & 13.60 & 1.45E+42 & 1962.101 &    57.7 & 15.90 & 1.75E+41 & 1964.052 &   770.3 & 18.50 & 1.59E+40 \\
1961.698 &   -89.5 & 13.97 & 1.03E+42 & 1962.139 &    71.6 & 15.80 & 1.91E+41 & 1964.090 &   784.2 & 18.35 & 1.83E+40 \\
1961.704 &   -87.3 & 13.95 & 1.05E+42 & 1962.148 &    74.9 & 16.30 & 1.21E+41 & 1964.186 &   819.3 & 18.40 & 1.75E+40 \\
1961.707 &   -86.2 & 14.02 & 9.86E+41 & 1962.172 &    83.6 & 16.50 & 1.00E+41 & 1964.534 &   946.4 & 18.15 & 2.20E+40 \\
1961.708 &   -85.8 & 14.10 & 9.16E+41 & 1962.184 &    88.0 & 16.70 & 8.36E+40 & 1964.610 &   974.1 & 18.15 & 2.20E+40 \\
1961.710 &   -85.1 & 14.00 & 1.00E+42 & 1962.241 &   108.8 & 16.70 & 8.36E+40 & 1964.625 &   979.6 & 17.70 & 3.33E+40 \\
1961.712 &   -84.4 & 14.06 & 9.51E+41 & 1962.518 &   210.0 & 16.70 & 8.36E+40 & 1964.682 &  1000.4 & 17.60 & 3.65E+40 \\
1961.715 &   -83.3 & 13.86 & 1.14E+42 & 1962.564 &   226.8 & 16.70 & 8.36E+40 & 1964.710 &  1010.6 & 17.70 & 3.33E+40 \\
1961.718 &   -82.2 & 14.04 & 9.68E+41 & 1962.567 &   227.9 & 16.69 & 8.43E+40 & 1964.745 &  1023.4 & 17.90 & 2.77E+40 \\
1961.721 &   -81.1 & 13.90 & 1.10E+42 & 1962.580 &   232.7 & 16.68 & 8.51E+40 & 1964.786 &  1038.4 & 17.90 & 2.77E+40 \\
1961.750 &   -70.5 & 14.08 & 9.33E+41 & 1962.600 &   240.0 & 16.70 & 8.36E+40 & 1964.868 &  1068.4 & 17.70 & 3.33E+40 \\
1961.770 &   -63.2 & 14.05 & 9.59E+41 & 1962.650 &   258.2 & 16.70 & 8.36E+40 & 1964.901 &  1080.4 & 18.00 & 2.52E+40 \\
1961.773 &   -62.1 & 14.10 & 9.16E+41 & 1962.660 &   261.9 & 16.80 & 7.62E+40 & 1964.902 &  1080.8 & 18.00 & 2.52E+40 \\
1961.775 &   -61.4 & 14.00 & 1.00E+42 & 1962.665 &   263.7 & 16.72 & 8.20E+40 & 1964.915 &  1085.5 & 17.90 & 2.77E+40 \\
1961.776 &   -61.0 & 13.80 & 1.21E+42 & 1962.671 &   265.9 & 16.80 & 7.62E+40 & 1964.975 &  1107.4 & 18.10 & 2.30E+40 \\
1961.786 &   -57.3 & 14.00 & 1.00E+42 & 1962.682 &   269.9 & 16.70 & 8.36E+40 & 1965.016 &  1122.4 & 18.10 & 2.30E+40 \\
1961.825 &   -43.1 & 13.86 & 1.14E+42 & 1962.683 &   270.3 & 16.74 & 8.05E+40 & 1965.156 &  1173.5 & 18.90 & 1.10E+40 \\
1961.827 &   -42.4 & 14.25 & 7.98E+41 & 1962.726 &   286.0 & 16.90 & 6.95E+40 & 1965.175 &  1180.5 & 18.50 & 1.59E+40 \\
1961.828 &   -42.0 & 14.30 & 7.62E+41 & 1962.742 &   291.8 & 17.15 & 5.52E+40 & 1965.178 &  1181.6 & 18.10 & 2.30E+40 \\
1961.829 &   -41.6 & 14.40 & 6.95E+41 & 1962.748 &   294.0 & 17.30 & 4.81E+40 & 1965.569 &  1324.4 & 18.40 & 1.75E+40 \\
1961.830 &   -41.3 & 14.30 & 7.62E+41 & 1962.750 &   294.8 & 16.98 & 6.46E+40 & 1965.600 &  1335.7 & 18.40 & 1.75E+40 \\
1961.838 &   -38.4 & 14.10 & 9.16E+41 & 1962.751 &   295.1 & 17.15 & 5.52E+40 & 1965.732 &  1383.9 & 18.40 & 1.75E+40 \\
1961.847 &   -35.1 & 14.20 & 8.36E+41 & 1962.759 &   298.0 & 17.20 & 5.27E+40 & 1965.759 &  1393.8 & 18.40 & 1.75E+40 \\
1961.852 &   -33.2 & 14.00 & 1.00E+42 & 1962.775 &   303.9 & 17.38 & 4.47E+40 & 1968.729 &  2478.6 & 21.20 & 1.32E+39 \\
1961.863 &   -29.2 & 13.70 & 1.32E+42 & 1962.805 &   314.8 & 17.50 & 4.00E+40 & 1968.805 &  2506.3 & 21.00 & 1.59E+39 \\
\enddata
\tablecomments{The integrated emission during the 970 days from date 1961.539
  to 1964.186 is $2.4 \times 10^{49}$ erg} \lTab{61vlite}
\end{deluxetable*}

\section{Presupernova Models}
\lSect{presn}

\subsection{Code Physics}
\lSect{physics}

The KEPLER code \citep{Wea78,Woo02} has been used to model PPISN many
times \citep[e.g.,][]{Woo17,Woo21} and the code physics used here is
the same as in those previous studies, except where noted in this
section. Standard mass loss prescriptions \citep{Nie90,Wel99} and
opacities were assumed, but the mass-loss rates were multiplied by an
adjustable parameter. This parameter, which was always less than one,
even after including scaling for a possibly sub-solar metallicity, was
adjusted to give pre-SN hydrogen envelope masses in the desired
range. More recent studies of mass loss in very massive stars suggest
that the mass-loss rates of \citet{Nie90} might be overestimates by a
factor of order three for such massive stars due the neglect of the
effects of clumping \citep[e.g.,][]{Smi14}. The multipliers we use are
thus moderate compared to what would be needed in more modern
treatments. Revising the mass-loss rate would also have the salutary
effect of making PPISN more easily achievable in higher metallicity
environments. For now though, the mass-loss rate multiplier is just a
reasonable artifice to achieve a desired pre-SN envelope mass. This
multiplier is also sensitive to the radius of the star, especially
during helium burning and was closer to unity for smaller radii. The
radius, in turn, is sensitive to convection physics and surface
boundary conditions (see below).

A composition appropriate to one-third solar metallicity was
adopted. That is, the initial abundances of hydrogen and helium were
0.719 and 0.276 by mass fraction and those of heavier
species were those of \citet{Lod03} multiplied by one third. For
example. the mass fractions of carbon, nitrogen, oxygen and iron were
$7.8 \times 10^{-4}$, $2.7 \times 10^{-4}$, $2.3 \times 10^{-3}$, and
$4.2 \times 10^{-4}$. All calculations employed a large adaptive
nuclear reaction network of approximately 300 to 400 species from
hydrogen through krypton coupled directly to the stellar
structure. Silicon quasiequilibrium was {\sl not} assumed at any time, up to
and including iron core collapse. A large network was necessary to
accurately follow weak interactions after carbon burning, including
those that happen during silicon burning, but it was also necessary to
avoid the quasiequilibrium approximation because of the long cool
inter-pulse phases, often below 10$^9$K, in many of the models where
silicon had already begun to burn at the center. This is the same
approach used by \citet{Woo17} and caused no difficulty here.

Standard values were used for the nuclear reaction rates, except for
the triple-alpha (3$\alpha$) reaction of helium burning.  It is now
understood that the presence of carbon in the pre-SN star
weakens the pair instability \citep{Far19,Far20,Woo21} and changes its
outcome. Carbon burning plus neutrino losses is never exoergic in the
centers of such massive stars, but carbon shell burning contributes
appreciable energy off-center both prior to and during the pulses.
This slows the collapse and can, especially in lower mass models, lead
to oxygen even igniting stably for a time at the star's center,
halting the collapse completely.  Eventually the carbon shell burns
out and the pulses commence, but fuel and momentum have been
diminished. This not only weakens the explosion, but makes the
properties of all but the first pulse sensitive to the uncertain
mixing that occurs during the interpulse periods. Mixing can bring new
carbon down to a mass coordinate of $\sim10$ \Msun \ where its burning
again slows the collapse. Mixing length convection was left on in the
remaining core during the interpulse periods, but not in the presence
of shocks or in the ejecta.

To keep the number of parameters manageable, only two choices of
initial carbon abundance were used, with the values determined by a
variable 3$\alpha$ rate. One, nominally the ``low carbon'' case, used
the same reaction rates as \citep{Woo17}, which corresponds to a
choice of $f_{3\alpha}$ = 1, $f_{\rm Buch}$ = 1.2 in Table 1 of
\citet{Woo21}. Another, the ``high-carbon'' case, used $f_{3\alpha} =
1.5$ and $f_{\rm Buch}$ = 1.2, that is, the $3 \alpha$, and only the
$3 \alpha$ rate was increased. This had the effect of approximately
doubling the carbon mass fraction after helium burning and gives
results similar to the ``1/1.35'' case in Table 1 of \citet{Woo21}.
As we shall see, this larger carbon abundance has the effect of
increasing the helium core masses required for a given duration of
pulsing activity by about 20\% and generally weakening the explosion.
The range explored probably spans the experimentally expected error
bars for the $^{12}$C($\alpha,\gamma)^{16}$O and 3$\alpha$ reaction
rates.

Also important and uncertain is the treatment of convection in the
hydrogen envelope of the progenitor star. Red and blue supergiant
envelopes with the same mass atop of the same helium and heavy element
core will have different interactions when traversed by a shock
wave. Stars with smaller radii will, initially at least, give fainter
SNe.  SN~1987A is the most familiar example of this.  Unfortunately,
convection in the envelopes of very massive, radiation-dominated stars
is not well understood.  Crudely, the convection is represented using
``mixing length'' theory, where the characteristic length for
convective energy dissipation is some factor, $\alpha$, times the
pressure scale height. Semiconvection, convective overshoot mixing,
convective dredge up, and rotational mixing are additional
complications.  Historically, KEPLER has used a mixing length
parameter, $\alpha$ = 1.  With modern opacities and this value of
$\alpha$, KEPLER has tended to give SNe~II-P progenitors with radii
that were too large by about a factor of two compared with what was
needed to properly replicate the light curve \citep{Des11,Des13}. As
those authors noted, increasing $\alpha$ to 2 or 3 reduced the pre-SN
radius by the necessary amount \citep[see also][]{Pax18}. A smaller
radius during helium burning would also help to bring KEPLER models
into better agreement with those of the Geneva group
\citep[e.g.,][]{Sch93}. Additionally, recent three-dimensional
simulations \citep{Gol21} have confirmed that, for red supergiant
envelopes, a value of $\alpha$ of 2 or 3 is a better choice, at least
in the MESA code. We thus explored two sets of models with larger
$\alpha$, as well as our more traditional case of $\alpha =1$.

Surface boundary conditions also matter. \citet{San15} have discussed
the radius inflation, up to a factor of 40, that occurs in their
models for stars with main sequence mass over 40 \Msun. The fact that
the luminosity is effectively super-Eddington in regions with high
opacity near the surface leads to convection, pulsations, radius
expansion, and density inversions. The authors speculated that the
instabilities they observed might be related to luminous blue variables
\citep[see also][]{Jia18}. We see the same effects in our models here,
all of which are over 90 \Msun. With the customary low surface
boundary pressure ($P_{\rm bound} = 100$ dyne cm$^{-2}$), pre-SN
radii exceeding $1.5 \times 10^{14}$ cm are common as well as density
inversions of up to an order of magnitude in the outer solar mass.

\citet{San15} do not specify what the actual radius should be. It is
probably variable, but the absence of observed red supergiants with
very high mass, and the association of some SNe~IIn with LBV-like
progenitors \citep{Smi14} suggests the probability of smaller
radii. To explore this possibility, we used a greatly increased
surface boundary pressure, 5000 dyne cm$^{-2}$, in a subset of models
to at least partly suppress radius expansion. This is a problem worth
further investigation, but for now the goal is just to bracket the
expected radii of SN 1961V progenitors, even if the physics is not
well understood.

\begin{figure}
\includegraphics[width=\columnwidth]{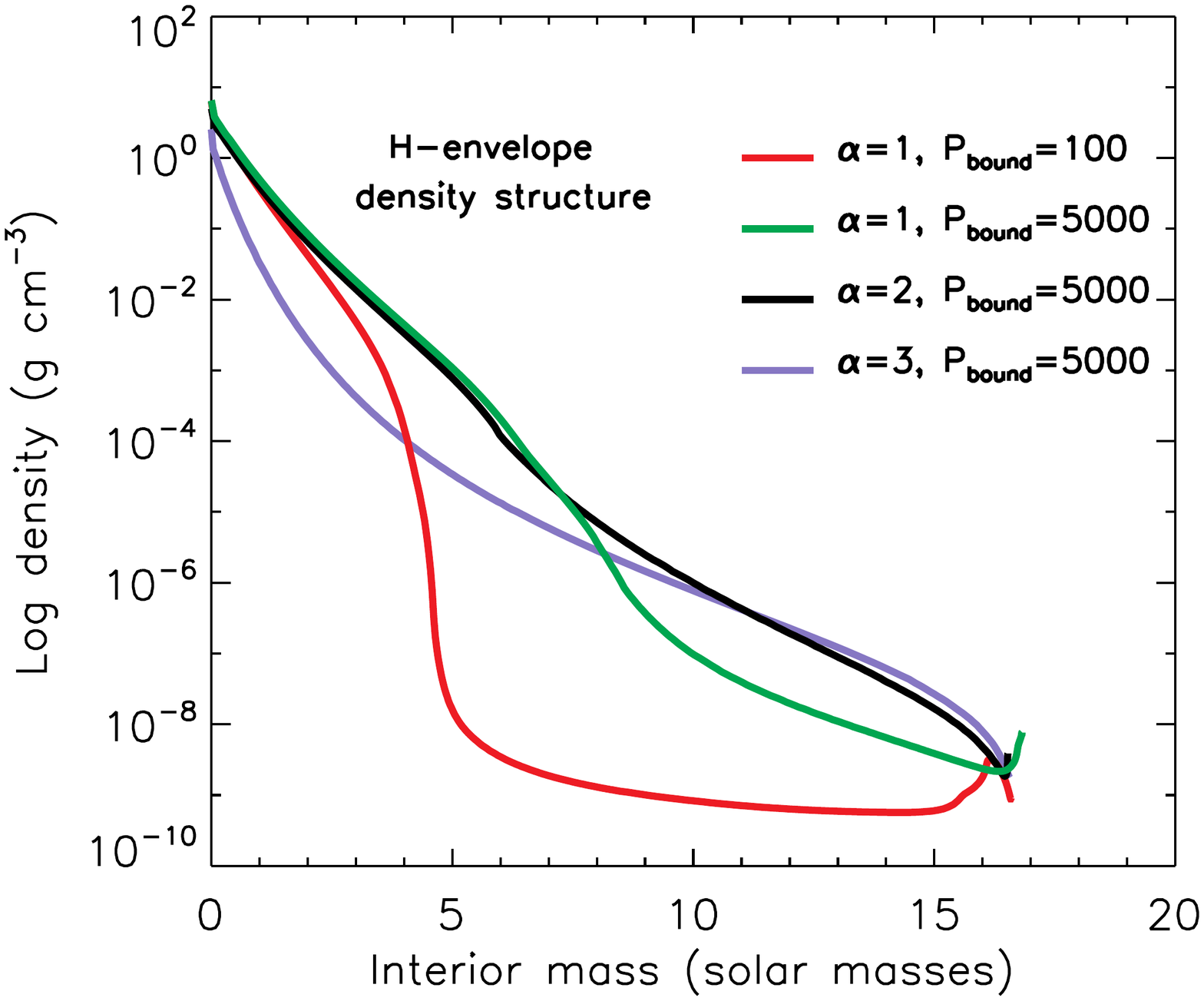}
\includegraphics[width=\columnwidth]{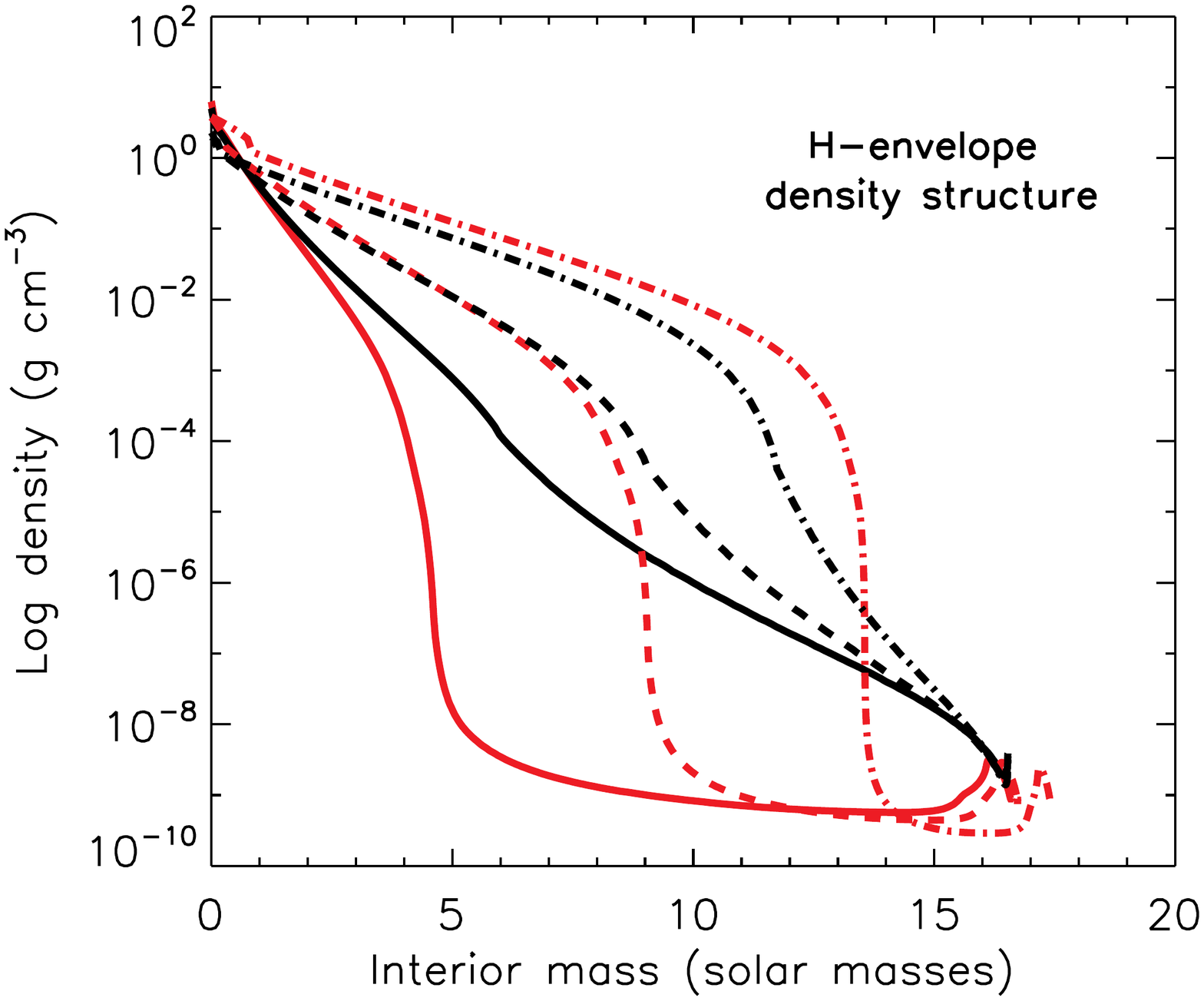}
\caption{Density structure of the hydrogen-rich envelope of several
  models with nearly the same final helium core and envelope
  masses. (top:) The structure is evaluated at the onset of
  pulsational activity ($T_c =$ 1.2 GK) for four models with helium
  core masses (not plotted) near 46.5 \Msun \ and hydrogen envelopes
  near 17 \Msun, but different values of mixing length parameters
  ($\alpha$) and surface boundary pressures (P$_{\rm bound}$). Zero on
  the mass scale is the base of the hydrogen envelope. The models are
  TH102p25, TH102p45*, M105p6, and M106p75 in \Tab{tmodels}. Higher
  surface boundary pressure and greater mixing lengths lead to denser,
  more compact envelopes with smaller radii. (bottom:) The density
  structure for Models TH102p25 (red lines, $\alpha = 1$,P$_{\rm
    bound}$ = 100 dyne cm$^{-2}$) and M105p6 (black lines, $\alpha =
  2$,P$_{\rm bound}$ = 5000 dyne cm$^{-2}$) evolve considerably
  following helium depletion (dash-dotted lines) and pre-carbon
  ignition ($T_c =$ 0.5 GK; dashed lines).  The solid lines show the
  structure at carbon depletion ($T_c =$ 1.2 GK) as in the top
  frame. In these two models, a surface convection zone eats deeper
  into the envelope as the core contracts moving more matter to larger
  radii. Note surface density inversions in Model
  TH102p25. \lFig{dn}}.
\end{figure}

To illustrate the sensitivity, four models were calculated that were
constrained, by varying initial mass and mass-loss rate, to give
approximately the same final helium core and hydrogen envelope masses.
These models used different overshoot parameters
$\alpha$ = 1, 2, and 3, and surface boundary pressures. These are
models TH102p25, TH102p45*, M105p6, and M106p75 in
\Tab{tmodels}. \Fig{dn} shows the hydrogen envelope structure late
in their evolution. The dominant effect comes from increasing the
surface boundary pressure, which has the effect of suppressing the low
temperature, high opacity regions near the surface. This helps
alleviate the density inversion and leads to a more compact
structure. The mixing length also has an important secondary
effect, larger values making the envelope more compact.

\Fig{dn} (bottom panel) also shows that, for at least two of these
models, TH102p25 and M105p6, much of the difference in pre-SN
structure develops after helium depletion ($Y_c < 0.01$) as a deep
surface convective zone eats into the envelope. This is a brief period
in the life of the star. From helium depletion to explosion is only
about 10,000 years, and from $T_c =$ 0.5 GK to explosion it is
1600 years. Unless the SN progenitor itself is observed, observers are
unlikely to catch a star during this phase of its life, but the
changes are significant for the pre-SN structure.

Fortunately, since we are only interested in the main light curve and
not shock break out, the initial radius of the star doesn't matter
much except for the first pulse, i.e., the beginning of long complex
events and PISN. It also matters, to a lesser extent, for the early
CSM interaction which depends on how much matter is accelerated to
high velocity by shock break out. The first pulse ejects the residual
hydrogen envelope and what happens afterward depends more on the
energy of that first pulse and the waiting time until the next one
than the initial radius.  Shock break out in very massive Type II SNe
like those studied by \citet{Kas11} may need reexamining.

\subsection{The Allowed Mass Range}

The properties of a Type II PPISN progenitor are chiefly determined
by: a) the mass of its helium core; b) the mass fraction of
carbon in that core at the time of central carbon ignition; and c) the
mass and radius of its hydrogen envelope (see \Tab{pisntab} and
\Tab{tmodels}). The energy of the explosion, the duration of pulsing
activity, and the intervals between pulses are all set by the helium
core mass and carbon fraction.  The light curve and spectrum are
additionally sensitive to the hydrogen envelope mass and its
structure. Secondary parameters are the metallicity, which affects the
mass loss, opacity, and the strength of the hydrogen burning shell;
the density profile within the hydrogen envelope, which also affects
the light curve; the opacity itself, which affects the radius; and the
treatment of convection in the model, both before and between the
pulses.  Neglecting rotation, as we do here, adds additional
uncertainty as does the one-dimension treatment of the explosion
hydrodynamics.

That said, any two stellar evolution calculations that give the same
helium core mass, carbon mass fraction, and hydrogen envelope mass
will produce qualitatively similar explosions. The initial mass,
metallicity, and mass loss rate matter chiefly by affecting these
three derived quantities.

Previous work on PPISNe allows us to narrow the range of He core
masses and other parameters for the special case of SN~1961V. For the
nuclear reaction rates and convection physics assumed in
\citet{Woo21}, having pulses that last at least a few weeks in order
to add structure to the $\sim$150 day light curve of SN 1961V, rules
out pre-SN helium cores less massive than $M_{\rm He} \ltaprx $ 40
\Msun \ and zero age main sequence masses less than $M_{\rm ZAMS}
\ltaprx 95$ \Msun, although this corresponding initial mass depends
somewhat on the adopted mass-loss rates. If we look for explosions
with pulsational activity comparable to the 200 day duration of SN
1961V, then core masses $47 \pm 1$ \Msun ($M_{\rm ZAMS} \approx 100$
\Msun) are preferred.  See also Figs 15 - 17 of \citet{Woo17}. These
choices are sensitive to the carbon abundance in the pre-SN star.

The largest helium core masses of interest are for cases where the
pulsational activity goes on much longer than SN 1961V itself.  One
possibility (\Sect{long}) is that only the first pulse or two make SN
1961V, but the star survives until much later, pulsing again just
before dying, but perhaps not producing a bright optical display at
that time. The gross upper limit for that case is then the boundary
between pair instability and pulsational pair instability SN which is
M$_{\rm He} \approx 65$ \Msun \ and M$_{\rm ZAMS} \approx 140$ \Msun
\ for the low carbon case and M$_{\rm He} \approx 70$ \Msun \ and
M$_{\rm ZAMS} \approx 150$ \Msun \ for the high-carbon case. Most of
these very high mass stars will have initial flashes that are too
energetic for SN 1961V. In practice, it turns out that mass limits
only about 20\% greater than for the single event model are
appropriate. Even given the uncertainty in carbon abundance, the
helium core mass for all successful PPISN models for SN 1961V is
probably between 45 and 55 \Msun, corresponding to a zero age main
sequence mass between 100 and 130 \Msun. If one includes the unlikely
PISN models of \Sect{pisn}, these limits are raised to 75 \Msun \ and
160 \Msun \ respectively, but these models are unlikely explanations.

Knowing the pre-SN helium core mass tightly constrains the
pre-SN luminosity of the star and the energetics of the
explosion. In KEPLER, the luminosity at the onset of carbon burning
for stars with helium cores in the mass range 45 to 55 \Msun \ is
limited to 1 to 1.3 $\times 10^{40}$ erg s$^{-1}$. In some cases, the
luminosity actually decreases slightly just prior to explosion in
response to the contracting core, so these are upper limits. We shall
find that the total kinetic energy of the explosion for stars in this
mass range is near $7 \times 10^{50}$ erg for the low carbon case and
$4 \times 10^{50}$ erg for the high-carbon case.  Most of the mass
ejected is residual helium-hydrogen envelope, which could in principle
range from 0 to (M$_{\rm ZAMS}$ - M$_{\rm He}$) $\approx 50$
\Msun. Very small and large vales are unlikely though since mass loss
is not negligible for a reasonable choice of metallicity, and a long
plateau is not possible with too little envelope. Taking $M_{\rm ej}
\approx$ 10 - 15 \Msun \ as representative of the ejected mass, a
typical velocity is then $(2 E_{\rm exp}/M_{\rm ej})^{1/2} \approx$
1600 - 2600 km s$^{-1}$, though one expects variation with higher
velocities near the outer edge and lower ones deep inside. That this
ballpark number agrees with what is observed at late times for the
speed of the bulk of the SN 1961V ejecta \citep{Bra71} is encouraging.

These estimates suggest a focus on models with masses M$_{\rm ZAMS}
\approx 95$ to 115 \Msun\ for the low carbon case and M$_{\rm ZAMS}
\approx 115$ to 135 \Msun \ for the high-carbon case. A few more
massive cases were considered in \Sect{pisn}. A metallicity of 1/3
solar was chosen for all models, consistent with estimates for the
vicinity of SN 1961V, and the mass-loss rate was adjusted to give
approximately desired envelope mass.  Again, somewhat lower mass-loss
rates (i.e. even lower than just scaling standard rates for
metallicity) are justified due to recent downward revisions in stellar
wind mass-loss rate prescriptions \citep{Smi14}. In general, the
successful explosions had lost most of their hydrogen envelope, but
still retained an appreciable amount.

The mass ranges and envelope masses motivated by these considerations
then define the models selected for study (\Tab{pisntab} and
\Tab{tmodels}).  The models are named according to the choices of
initial mass, mass-loss rate, mixing length parameter, and boundary
pressure. Not all combinations were explored. All models used the same
one-third solar metallicity composition. The ``TH'' models \citep[here
  ``TH'' is for third solar metallicity to distinguish them from the
  ``T'', tenth solar metallicity models, of][]{Woo17}, had $\alpha =1 $
and P$_{\rm bound}$ = 100 dyne cm$^{-2}$. The ``M'' models (``M'' is
for ``mixing length'') used a mixing length parameter $\alpha$ = 3 and
a boundary pressure of 5000 dyne cm$^{-2}$. The ``A'' models used a
mixing length parameter $\alpha$ = 2, a boundary pressure of 5000 dyne
cm$^{-2}$, and a large $3 \alpha$ rate as described above so as to give
a high carbon abundance after helium burning. Models were further named
for their initial mass and mass-loss rate reduction, e.g., Model T102p25 had an
initial mass of 102 \Msun\ and a mass-loss rate 25\% of the standard
value. Sometimes the ``pxx'' subscript is dropped in the plots where no
ambiguity arises, e.g., most of the TH models used p25 for mass loss.

\subsection{Presupernova Evolution}

The stars here all lived approximately 2.7 million years as bright,
hot main sequence stars with effective temperatures near 50,000 K and
luminosities 0.6 (90 \Msun) to 1.0 (140 \Msun) $\times 10^{40}$ erg
s$^{-1}$. This was followed by approximately 300,000 years as helium
burning giant stars with luminosities from 0.8 to 1.4 $\times 10^{40}$
erg s$^{-1}$. The main sequence evolution is straightforward and has
been studied many times \citep[e.g.][]{Sch93}. The radius during
helium burning is less certain and dependent on convection and surface
boundary pressure (\Sect{physics}). For $\alpha$ = 1, the models all
developed deep surface convection zones at the end characterized by large radii
and density inversions. That is they experienced radius
inflation and became cool supergiants. Halfway through helium burning,
typical effective temperatures were $\approx 4500$ K.

For models with $\alpha$= 2 or 3 and high surface pressures, the
luminosities were nearly the same, but the radii were smaller by a
factor of a few. Effective temperatures were thus 7500 to 8500 K.  In
this case, both as helium burning stars and SN progenitors, the stars
would most likely have appeared as F-supergiants \citep[see
  also][]{Goo89} with a relatively small bolometric correction. This
is consistent with expectations for eruptive LBV-like stars
\citep{Smi04}.

Since the progenitor of SN 1961V was repeatedly observed starting in
1937 (\Tab{61vlite}), the last 25 years of the models' lives are of
special interest. During this time, all models were contracting from
carbon ignition at about 0.75 GK to pair instability when
oxygen ignited around GK. Most of the star's luminosity during this
time, both in photons and neutrinos was derived from this contraction
and not nuclear fusion. Unlike lower mass SNe, there is no extended
period of carbon (centuries) or oxygen (months) burning. During this
contraction, no convection occurred inside the helium core until the
last week. Thus there was very limited time for acoustic transport to
the surface of a large energy flux that might drive mass loss
\citep{Shi14}. In the last few months, the hydrogen burning shell was
super-heated by the contraction of the helium core and generated a
power that, were it communicated to the surface, would have been
super-Eddington, but there was insufficient time even for convection
to do so. The star expanded slightly in response. The star thus often
died in a state of thermal disequilibrium with the power emerging from
the helium core exceeding what was being emitted from the surface.

\section{Pair-Instability Models}
\lSect{pisn}

Stars with initial mass above 80 \Msun, the (smallest) lower limit
on progenitor mass imposed on SN~1961V by \citet{Koc11}, are not
expected to explode by neutrino transport or to leave a neutron star
\citep[e.g.][]{Heg02,Suk16,Rah22}. They either blow up completely as
pair-instability supernovae (PISN), leave black hole remnants
following an epoch of violent nuclear-fueled pulsations (PPISNe),
or collapse directly to black holes without mass ejection or bright
transient emission.

A PISN explanation is possible for SN 1961V, but not very
likely. Models with bolometric light curves having the same
approximate duration and luminosity as the major event of SN 1961V
have been published. See Model R150 and Fig. 8 of \citet{Kas11} and
Models P150 and P175 and Fig. 13 of \citet{Gil17}. If the entire star
explodes in one pulse though, a very oxygen-rich SN remnant would
result that doesn't seem to have been observed \citep{Goo89}.  The
total kinetic energy of the event would exceed $4 \times 10^{51}$ erg,
and much of the hydrogen-rich ejecta would move faster than 2000 km
s$^{-1}$, even if the envelope were massive. A PISN model, by itself,
would also leave unexplained the brightening to $m_{\rm pg}$ = 15.8
mag observed one year prior to the peak of the SN (\Tab{61vlite}) and
the years of emission after the peak. We shall find PPISN models
that fit SN 1961V better, and expect them to be more common in
nature. Still, PISN models are worth brief exploration, if only to
provide some interesting limits on the pre-SN mass and luminosity.

\begin{deluxetable*}{ccccccccccc} 
\tablecaption{Single Pulse Pair-Instability Supernova Models} 
\tablehead{ Mass & Mass Loss & $\alpha$ & M$_{\rm He}$ & X($^{12}$C) & M$_{\rm preSN}$ & R$_{\rm PreSN}$ & L$_{\rm PreSN}$  & M$_{\rm Ni}$ & KE$_{\rm eject}$ & E$_{\rm rad}$ \\
(\Msun) & Multiplier & & (\Msun) & &(\Msun)  & (10$^{13}$ cm)  & (10$^{40}$ erg s$^{-1}$) & (\Msun) & (10$^{50}$ erg) & (10$^{50}$ erg)}
\startdata
TH140p0    &  0   & 1 & 68.05 & 0.0791 &139.95 &  22 & 1.3 & 0.0280 & 45.4 & 2.41 \\
TH140p125 & .125 & 1 & 67.36 & 0.0675 &104.27 &  23 & 1.3 & 0.0450 & 54.8 & 1.95 \\
M140p25   & .25  & 2 & 66.53 & 0.0795 &107.05 &  7.5 & 1.2 & 0.0148 & 37.4 & 0.43 \\
M140p125  & .125 & 2 & 66.81 & 0.0776 &123.99 &  7.1 & 1.3 & 0.0316 & 47.6 & 0.46 \\
A155p2    & .20  & 2 & 74.96 & 0.126  &127.07 &  6.9 & 1.4 & 0.0102 & 57.1 & 0.48 \\
A160p2    & .20  & 2 & 77.18 & 0.127  &129.92 &  7.4 & 1.4 & 0.204  & 149  & 1.20 \\
\enddata
\tablecomments{The mass loss multiplier is applied in addition to the
  usual reduction for metallicity, Z$^{1/2}$.} \lTab{pisntab}
\end{deluxetable*}

Six PISN models (\Tab{pisntab}) were calculated that had the
approximate duration of SN 1961V. Each had a main sequence mass of 140
\Msun \ or more, a metallicity one-third solar, and a mass-loss rate
well below standard values expected for this metallicity.  Compared
with \citet{Kas11} and \citet{Gil17}, these models have the advantage
of a more realistic metallicity for SN~1961V, 1/3 $Z_{\odot}$ {\sl vs}
10$^{-4}$ $Z_{\odot}$, and possibly a more realistic radius, but they
have the disadvantage of using KEPLER to calculate their light curves
rather than the more capable transport codes SEDONA or STELLA. KEPLER
should be capable of providing approximate bolometric light
curves though. The pre-SN luminosities and radii in \Tab{pisntab} were
evaluated at carbon depletion, just a few weeks before pulsations
commenced, and are representative pre-SN properties. They evolve
very little during the last century of the star's life.  More massive
models than in \Tab{pisntab} would have been even more energetic
($\gtaprx10^{52}$ erg), produced brighter light curves, and larger
amounts of $^{56}$Ni ($> 0.1$ \Msun). The hydrogen envelope, even in
the absence of any mass loss, would move much faster than 2000 km
s$^{-1}$, and the $^{56}$Ni mass would exceed 0.2 \Msun \ producing
very bright tails on the light curves. These traits are incompatible
with observation, and we conclude that the progenitor of SN 1961V must
have had a bolometric luminosity less than $1.4 \times 10^{40}$ erg
s$^{-1}$ and an initial main-sequence mass less than 160 \Msun, even
if it was a PISN.

Unlike the pre-SN luminosity, the PISN light curves are
sensitive to the radius and structure of the envelope (\Sect{physics},
\Fig{ppilite}).  Four models, M140p25, M140p125, A155p2, and A160p2
employed a higher surface boundary pressure and larger mixing length
multiplier, and thus had reduced progenitor radii.  Effective
temperatures for the pre-SN stars were 7500 - 8500 K for M140p25,
M140p125, A155p2 and A160p2 and 4500 K for TH140p125 and TH140p0.
Typically the BSG progenitor radius was about three times smaller
than for the red supergiant models, TH140p0 and TH140p125
(\Tab{pisntab}). This resulted in light curves that were briefer and
less luminous. In addition, Models A155p2 and A160p2 used reaction
rates that favored a larger carbon abundance after helium burning and
were thus weaker explosions.

\begin{figure}
\includegraphics[width=\columnwidth]{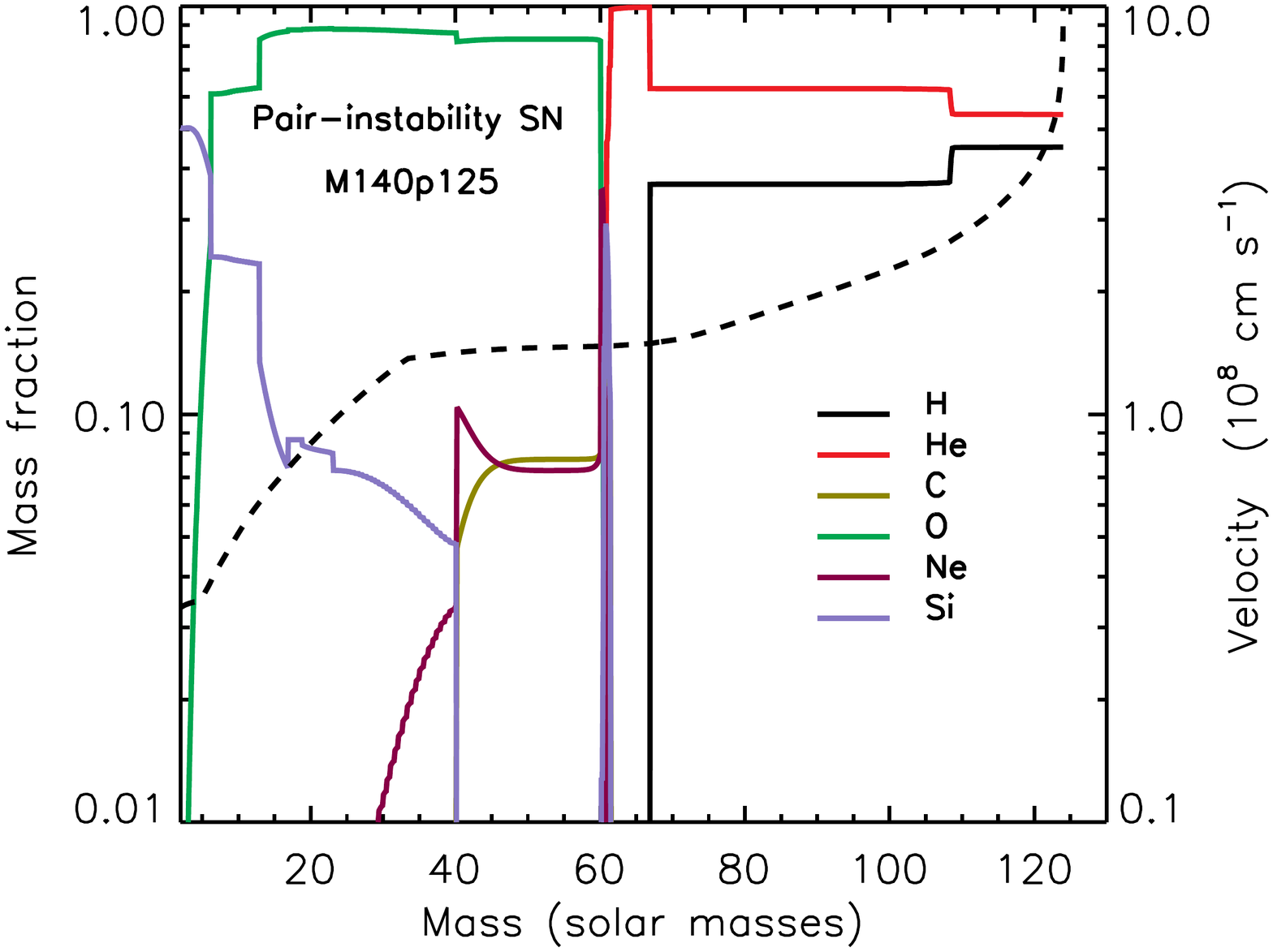}
\includegraphics[width=\columnwidth]{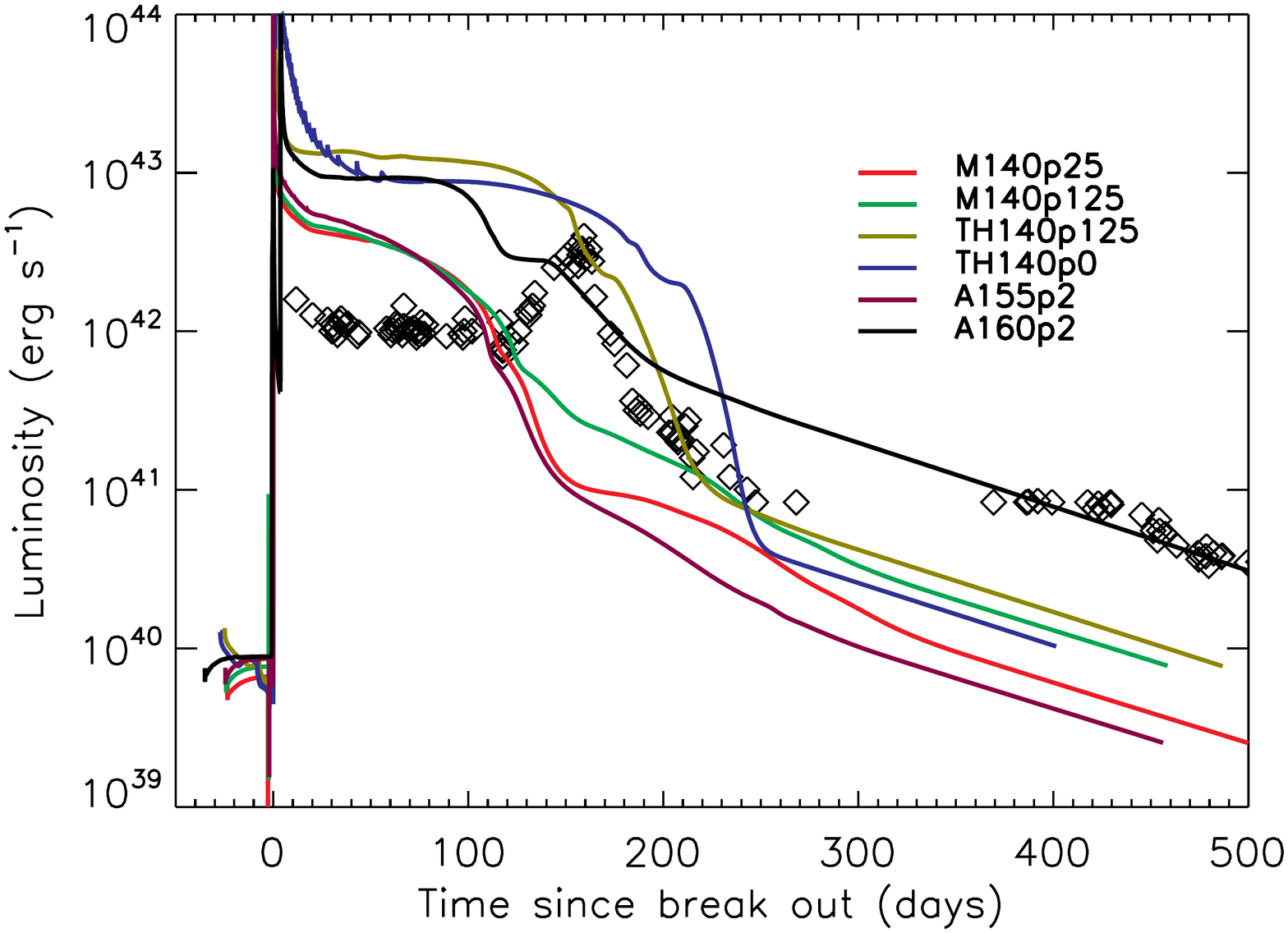}
\caption{(top:) Ejected composition and terminal velocity for Model
  M140p125 and (bottom:) approximate bolometric light curves for the
  six PISN models in \Tab{pisntab}. The velocity and composition of
  Model M140p125 are typical of the five models that have some
  similarity to SN 1961V, although the hydrogenic velocity is 
  slower for Model TH140p0 which had no pre-SN wind mass
  loss. Models M140p25, M140p125, A155p2, and A160p2 were blue
  supergiants and had a more compact structure when they
  exploded. Models TH140p0 and TH140p125 were red supergiants with
  very low density hydrogen envelopes. Tails due to the decay of
  $^{56}$Co are apparent. None of the models resembles closely the
  observations of SN 1961V, which is why we favor pulsational pair
    models for SN~1961V as opposed to a true PISN.\lFig{ppilite}}
\end{figure}

All PISN models produced small, but consequential amounts of $^{56}$Ni
(\Tab{pisntab}). The $^{56}$Ni mass is sensitive to the temperature
reached in the pulse and hence to the mass of the helium core and
carbon mass fraction. It would be larger for more massive helium cores
or less carbon in the pre-SN star. Model A160p2 is an upper
limit where even the luminosity on the tail of SN 1961V is
overproduced by radioactivity. The other models have durations,
hydrogen velocities, and luminosities crudely in agreement with SN
1961V, but with substantial disparities. The red giant models are too
bright, owing to their large initial radii and high explosion energy,
and can be ruled out. The bluer models are still too bright and
too brief, though the latter could partly be an artifact of using KEPLER with
just electron scattering opacity plus an additive constant. A
recalculation of Model R150 of \citet{Kas11} using just KEPLER gave a
shorter plateau, so this may be an issue worth revisiting.

But none of the light curves in \Fig{ppilite} show the late time
brightening observed when SN 1961V reached its peak, and they all
require relatively low mass-loss rates for a metallicity of 1/3 solar
in order to preserve a large mass of hydrogen envelope.  The models do
not explain the pre-explosive brightening one year before the observed
peak (\Tab{61vlite}), though this could be some sort of pre-SN LBV
outburst \citep{Goo89}. They also do not explain the enduring emission
seen for at least three years after the peak, though circumstellar
interaction could be invoked \citep{Smi11c}. The models predict a
large mass of oxygen (roughly 45 \Msun) and silicon (roughly 5 \Msun)
moving at very slow speed, $\sim1000$ km s$^{-1}$
(\Fig{ppilite}). Given the low level of radioactivity, perhaps these
very large masses might be difficult to detect, but by now they cover
a region roughly 0.1 pc in radius.

A PISN model is thus not completely ruled out for SN 1961V, but there are
PPISN models that look more promising and should occur more frequently

\section{Pulsational Pair-Instability Models}
\lSect{ppisn}

There are three subclasses of PPISN models that could plausibly
explain SN 1961V. In the first (\Sect{single}), the principle outburst
of SN 1961v was a single SN-like event consisting of multiple pulses
followed soon afterward by a final collapse to a black hole that
occurred while most of the ejecta was still optically thick. Most of
the pulses occurred early on, but the light curve continued to accumulate
energy well after core collapse. This model at least has
the advantage of providing a credible explosion mechanism for a very
massive progenitor, but other activity prior to or long after the main
event in 1961 must be attributed to variability of the pre-SN star
(before explosion) and shock interaction with the star's wind (after
the explosion).

\begin{deluxetable*}{cccccccccccc} 
\tablecaption{PULSATIONAL PAIR-INSTABILITY MODELS} 
\tablehead{ Mass & Mass Loss & $\alpha$ & M$_{\rm He}$ & X($^{12}$C) & M$_{\rm preSN}$& R$_{\rm PreSN}$ & M$_{\rm Fe}$ &  Duration & M$_{\rm final}$ & KE$_{\rm eject}$ & E$_{\rm rad}^{\, a}$ \\
(\Msun) & Multiplier  &  & (\Msun) & &(\Msun)  & (10$^{14}$ cm) & (\Msun) & (days) & (\Msun) & (10$^{50}$ erg) & (10$^{50}$ erg)}
\startdata
TH93p25 & .25 & 1 & 40.44 & 0.113  & 60.13 &  18 & 2.54 & 7.94  & 38.3 & 3.24 & 0.33 \\
TH95p25  & .25 & 1 & 42.68 & 0.109  & 60.49 &  18 & 3.23 & 20.5  & 39.4 & 5.26 & 0.53 \\
TH97p25  & .25 & 1 & 43.37 & 0.106  & 61.12 &  18 & 2.52 & 24.6  & 39.4 & 5.70 & 0.57 \\
TH99p25  & .25 & 1 & 44.79 & 0.100  & 61.59 &  18 & 2.86 & 44.2  & 40.0 & 6.27 & 0.82 \\
TH99p125 &.125 & 1 & 45.19 & 0.104  & 77.56 &  20 & 2.19 & 51.1  & 40.6 & 6.44 & 0.75 \\
TH100p25 & .25 & 1 & 45.50 & 0.0982 & 61.85 &  18 & 2.69 & 69.5  & 40.0 & 7.72 & 1.22 \\
TH101p25 & .25 & 1 & 45.60 & 0.100  & 62.43 &  18 & 2.52 & 61.8  & 40.2 & 7.53 & 1.15 \\
TH102p25 & .25 & 1 & 46.14 & 0.100  & 62.73 &  18 & 2.27 & 80.0  & 40.8 & 7.35 & 1.26 \\
TH102p45*$^b$ &.45 & 1 & 46.49 & 0.0991 & 62.93 & 7.9 & 2.71 & 56.3  & 42.1 & 8.35  & 0.89 \\
TH102p125 &.125 & 1 & 46.10 & 0.104  & 79.38 &  20 & 2.33 & 74.0  & 41.2 & 6.92 & 0.98 \\
TH103p25 & .25 & 1 & 46.68 & 0.0996 & 63.09 &  18 & 2.39 & 102   & 41.1 & 7.60 & 1.54 \\
TH104p25 & .25 & 1 & 47.01 & 0.0964 & 66.46 &  16 & 2.50 & 231   & 41.6 & 7.37 & 1.97 \\
TH105p25 & .25 & 1 & 48.84 & 0.0897 & 63.07 &  18 & 2.56 & 361   & 44.4 & 7.20 & 2.54 \\
TH106p25 & .25 & 1 & 47.34 & 0.0996 & 64.32 &  18 & 2.72 & 188   & 42.3 & 7.31 & 1.80 \\
TH107p25 & .25 & 1 & 50.03 & 0.0882 & 63.66 &  18 & 2.16 & 567   & 44.2 & 8.20 & 3.14 \\
TH108p25 & .25 & 1 & 49.71 & 0.0916 & 64.38 &  18 & 2.20 & 440   & 44.3 & 7.64 & 2.94 \\
TH109p25 & .25 & 1 & 50.05 & 0.0938 & 64.03 &  18 & 2.89 & 590   & 44.4 & 7.20 & 3.58 \\
TH110p25 & .25 & 1 & 50.49 & 0.0911 & 65.09 &  18 & 2.71 & 703   & 44.7 & 8.02 & 3.02 \\
TH111p25 & .25 & 1 & 51.36 & 0.0907 & 65.36 &  18 & 2.53 & 1050  & 46.9 & 7.04 & 3.92 \\
TH112p25 & .25 & 1 & 51.74 & 0.0907 & 65.80 &  18 & 2.14 & 1300  & 47.2 & 7.26 & 3.99 \\
TH113p25 & .25 & 1 & 51.32 & 0.0935 & 66.47 &  18 & 2.38 & 872   & 45.7 & 7.56 & 3.91 \\
TH114p25 & .25 & 1 & 51.63 & 0.0915 & 70.18 &  16 & 1.93 & 14600 & 46.1 & 8.33 & 1.35 \\
TH115p25 & .25 & 1 & 52.12 & 0.0901 & 70.51 &  16 & 1.99 & 39100 & 45.9 & 9.70 & 1.77 \\
TH120p25 & .25 & 1 & 54.56 & 0.0899 & 72.40 &  16 & 2.42 & 12900 & 48.7 & 11.5 & 0.51 \\
          &     &   &       &        &       &     &      &       &      &      &      \\
M104p9    & 0.9 & 3 & 45.34 & 0.111  & 55.52 & 4.9 & 2.34 & 32.1  & 41.3 & 5.27 & 0.41 \\
M105p9    & 0.9 & 3 & 45.55 & 0.109  & 55.56 & 4.9 & 2.58 & 37.9  & 41.2 & 5.39 & 0.46 \\
M105p6    & 0.6 & 2 & 47.11 & 0.096  & 63.64 & 5.7 & 2.23 & 85.1  & 42.2 & 8.19 & 1.13 \\
M106p9    & 0.9 & 3 & 46.14 & 0.110  & 55.83 & 4.9 & 2.50 & 40.2  & 41.6 & 5.71 & 0.48 \\
M106p75   & 0.75& 3 & 46.90 & 0.107  & 63.01 & 5.0 & 2.74 & 69.2  & 42.7 & 5.80 & 0.61 \\
M107p9    & 0.9 & 3 & 46.72 & 0.104  & 55.67 & 4.9 & 2.46 & 78.4  & 41.8 & 5.31 & 0.69 \\
M108p85   & .85 & 3 & 47.53 & 0.108  & 58.44 & 5.0 & 2.66 & 83.1  & 42.6 & 6.32 & 0.86 \\
M109p85   & .85 & 3 & 48.05 & 0.107  & 58.55 & 5.0 & 2.68 & 110   & 43.8 & 5.75 & 0.85 \\
M110p85   & .85 & 3 & 48.53 & 0.106  & 58.61 & 5.0 & 2.27 & 160   & 43.9 & 5.83 & 1.10 \\
          &     &   &       &        &       &     &      &       &      &      &      \\
A119p70   & .70 & 2 & 53.01 & 0.174  & 61.49 & 6.7 & 2.38 & 19.8  & 49.7 & 4.17 & 0.27 \\
A120p70   & .70 & 2 & 53.75 & 0.172  & 60.57 & 6.7 & 2.40 & 29.7  & 50.4 & 4.34 & 0.37 \\
A122p65   & .65 & 2 & 54.99 & 0.172  & 65.01 & 6.8 & 2.20 & 50.7  & 50.7 & 4.24 & 0.43 \\
A123p65   & .65 & 2 & 55.15 & 0.172  & 65.15 & 6.9 & 2.39 & 149   & 51.2 & 4.15 & 0.73 \\
A124p65   & .65 & 2 & 55.73 & 0.171  & 65.02 & 6.9 & 2.15 & 257   & 51.3 & 4.23 & 0.74 \\ 
A122p70   & .70 & 2 & 54.87 & 0.173  & 60.92 & 6.6 & 2.13 & 115   & 50.1 & 4.14 & 0.57 \\
A123p70   & .70 & 2 & 55.01 & 0.171  & 61.20 & 6.5 & 2.17 & 135   & 51.0 & 3.90 & 0.67 \\
A124p70   & .70 & 2 & 55.48 & 0.167  & 60.91 & 6.5 & 2.17 & 258   & 51.1 & 3.86 & 1.08 \\
A122p75   & .75 & 2 & 55.14 & 0.158  & 57.48 & 5.3 & 2.05 & 683   & 50.8 & 3.29 & 0.68 \\
A123p75   & .75 & 2 & 54.94 & 0.169  & 57.40 & 5.3 & 2.33 & 192   & 51.1 & 3.07 & 0.74 \\
A124p75   & .75 & 2 & 55.08 & 0.165  & 57.48 & 5.4 & 2.08 & 306   & 50.2 & 3.48 & 0.82 \\
A125p65   & .65 & 2 & 56.30 & 0.171   & 65.64 & 6.8 & 2.32 & 221   & 52.6 & 3.98& 1.00 \\
A126p65   & .65 & 2 & 56.54 & 0.168  & 65.48 & 6.8 & 2.23 & 382   & 51.7 & 4.01 & 1.22 \\
\enddata
\tablecomments{$^a$ The radiated energy, $E_{\rm rad}$, only includes
  the emission during the first 1000 days of pulsing activity.\\$^b$
  Model TH102p45 and all the ``M'' and ``A'' models had a surface
  boundary pressure of 5000 dyne cm$^{-2}$. The rest of the TH models
  had a boundary pressure of 100 dyne cm$^{-2}$. } \lTab{tmodels}
\end{deluxetable*}

In the second case (\Sect{long}), SN 1961V itself resulted from one or two
initial pulses, but those pulses were part of a continuing multiplet
that resumed, after a long dormant period, decades or even centuries
later. This is the only scenario where anything resembling a normal
star, could still occupy the site of SN 1961V several years
later. Powered by Kelvin-Helmholtz contraction, not nuclear burning,
and perhaps supplemented by accretion from fallback, the star's
luminosity would remain near Eddington. That is, it would be close to
its pre-explosive value. The star could still be there
or it could be gone by now, replaced by a black hole of about 45
\Msun.

The third case (\Sect{multiple}) assumes that SN 1961V underwent a
series of explosions lasting several years. The core has long since
collapsed to a black hole, but only after multiple shell ejections and
SN-like outbursts. The best fit to the onset of SN 1961V in this case
is obtained not by fitting the onset of the SN to the first
pulse, but to the second or third pulse happening hundreds of days later. In
this scenario, bright events prior to SN 1961V itself must somehow
have been missed by observers at the time. Because the colliding
shells are not very geometrically thick and the optical depth is low,
rapid temporal variation is possible, but collisions at very late
time might have low optical efficiency. The energy that is radiated is
given by the differential speed of colliding shells, which may be much
less than their bulk spectroscopic speeds.

For a given carbon mass fraction, these three possibilities correspond
to narrow ranges of helium core mass. For the ``TH'' models in
\Tab{tmodels} the appropriate ranges for the three outcomes are 46 -
48 \Msun, 50 - 52 \Msun, and 48 - 50 \Msun. No matter which subclass
or carbon abundance is invoked, the total mass of the pre-SN star is
constrained to be approximately $65 \pm 5$ \Msun \ and a black hole of
about $45 \pm 5$ \Msun \ is the final remnant. The more massive black
holes in this range are favored by a large carbon abundance.

\subsection{Single Events}
\lSect{single}

For relatively low-mass helium cores, pulsational activity is confined
to the first 100 days \citep{Woo17}. ``Single events'' here really
means a rapid succession of pulses that are grouped closely together
while the SN is still optically thick (e.g. \Fig{pulses}), producing a
single observed bright SN, albeit with a structured light curve.
Depending on the timing and energies of these pulses, collisions
between ejected shells may continue well after the black hole forms,
but these collisions do not persist to sufficiently late times to
explain the many years that SN 1961V maintained a luminosity in excess
of 10$^{40}$ erg s$^{-1}$.

\begin{figure}
\includegraphics[width=\columnwidth]{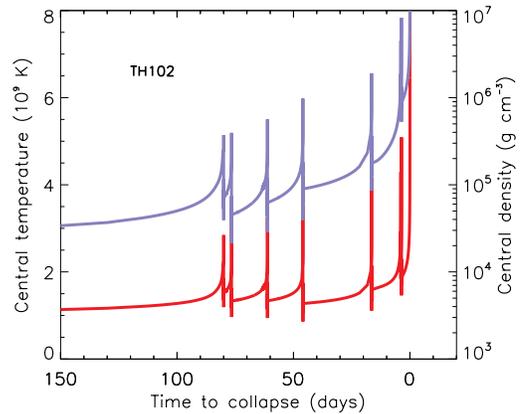}
\caption{Pulsing history of Model TH102p25, a representative example
  of the class of single event models. The central temperature and
  density are shown as a function of time to core collapse. Six well
  defined pulses happen in a space of 80 days. The last rise to a
  central temperature above 8 GK indicates collapse a black hole. No
  outward shock is generated by this final event.  \lFig{pulses}}
\end{figure}

\begin{figure}
\includegraphics[width=\columnwidth]{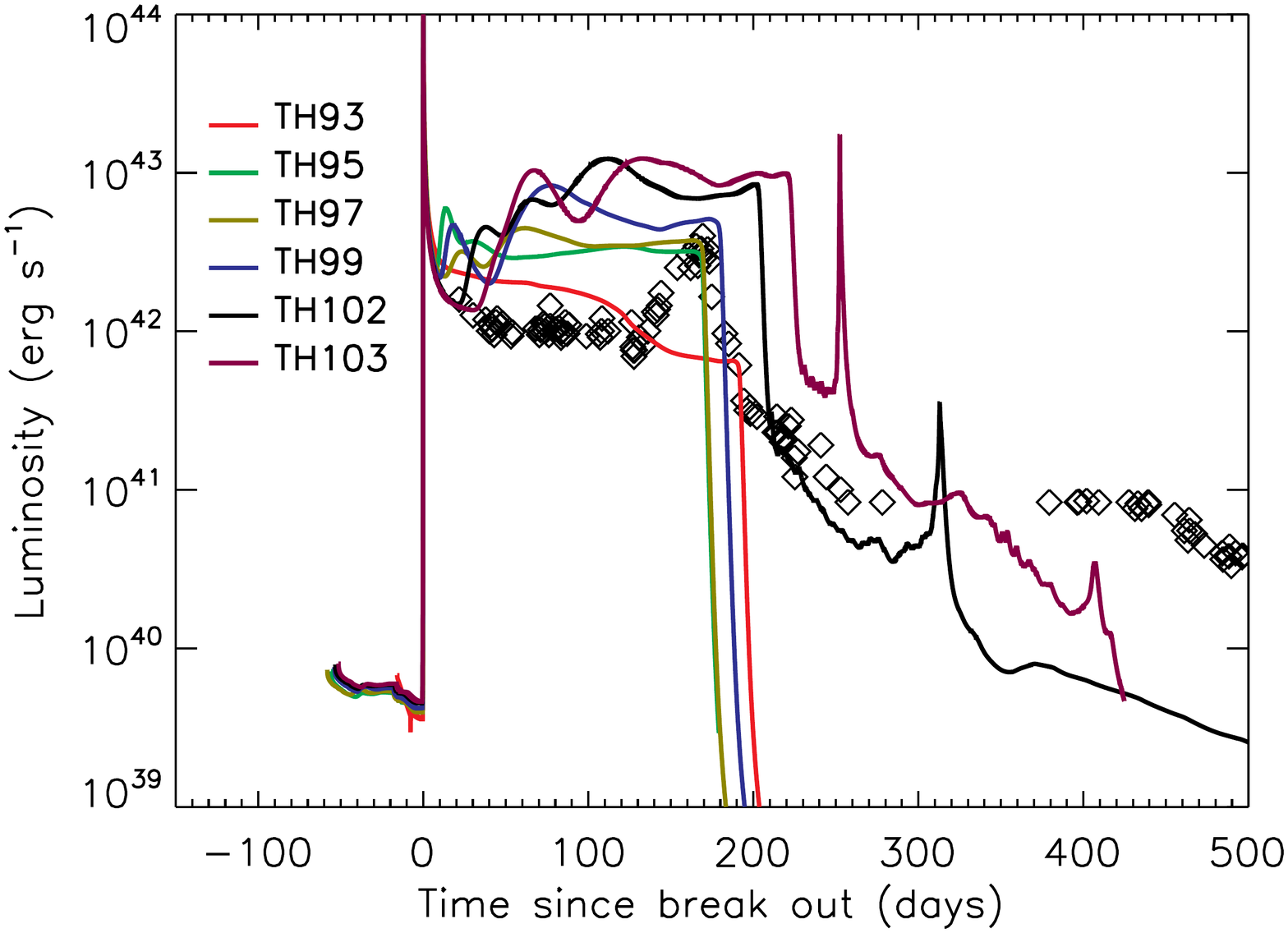}
\includegraphics[width=\columnwidth]{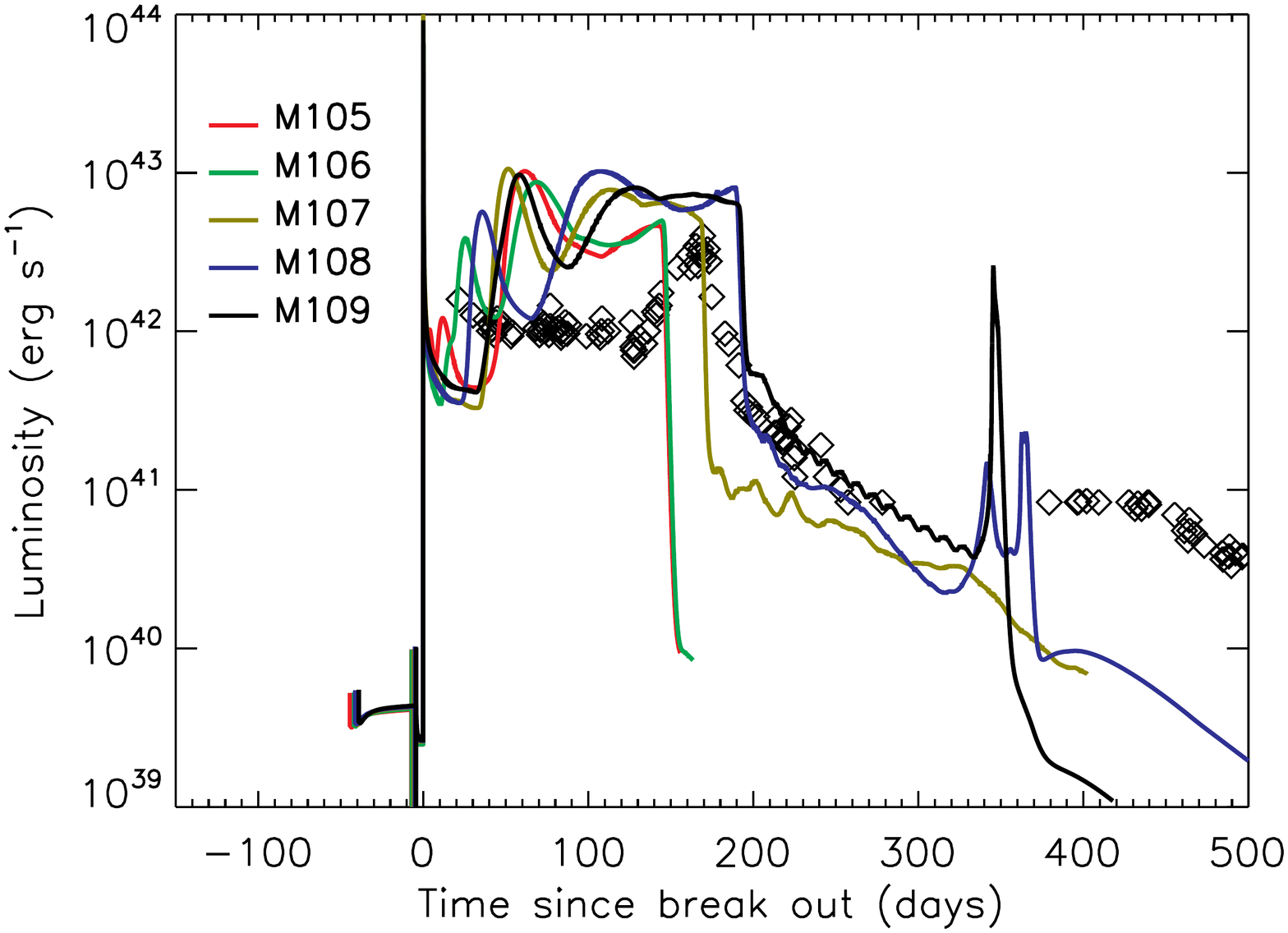}
\caption{(top:) Bolometric light curves of Models TH93p25, TH95p25,
  TH97p25, TH99p25, TH102p25, and TH103p25 (\Tab{tmodels}). In all
  cases, the duration of pulsing activity is shorter than that of the
  initial plateau in the light curve.  Multiple shocks from the pulses
  deposit energy in the deep interior that diffuses, in waves, to the
  photosphere. The greater the mass of the star the longer this
  pulsing persists. Later sharp structure in some of the light curve
  is introduced by fast moving thin shells that eventually catch up with
  slower ones after core collapse. (bottom:) Light curves for a
  similar set of models, M105p9, M106p9, M107p9, M108p85, and M109p85,
  with smaller radii and denser hydrogen envelopes. The initial masses
  are smaller because of the reduced mass-loss rates associated with
  the smaller radii. The light curves are similar to those for the
  ``TH'' models, except at early times when the smaller radius causes
  the emission to be fainter. \lFig{various}}
\end{figure}

\Fig{various} shows the light curves for six ``TH'' models and five
``M'' models in \Tab{tmodels} with total pulsational durations less
than 120 days (\Fig{pulses}). The ``TH'' models have identical physics
to \citet{Woo17} and ``low carbon'' abundances, i.e., a relatively
high rate for $^{12}$C($\alpha,\gamma)^{16}$O and low rate for
3$\alpha$ \citep{Woo21}. The final ejecta velocities and effective
emission temperatures are shown for these models in
\Fig{teff}. Models less massive than TH95 are disallowed by their low
velocities for envelope masses sufficient to maintain the continuity
and duration of the light curve.

Four phases of the light curve can be distinguished: 1) A bright,
brief, ultra-violet transient as the first shock breaks out; 2) A
plateau as the shock heated envelope is ejected by the first pulse and
recombines. This phase is fainter for models with smaller radii and longer
for more massive models, but not more than a few weeks. It is the
analogue of an ordinary Type II-P SN; 3) A series of broad peaks
generated by multiple pulses interacting with optically thick matter
ejected by previous pulses. The light diffuses ahead of the shock and
causes a significant brightening as it approaches the photosphere, but
each peak also has a ``tail'' that can be obscured by subsequent
pulses except for the last; and 4) Narrow spikes resulting from the
collision of geometrically thin shells. These shells are accumulations
of matter generated by pile up at both the forward and reverse shocks
associated with each pulse. The width of these spikes is
underestimated and the peak luminosity overestimated in this 1D study.

The durations in \Tab{tmodels}, which give the time between the
launching of the first pulse and iron core collapse, are only
approximations to the duration of the light curve. They can be
overestimates of the observed light curve duration, since it takes
up to two weeks for the first shock to traverse the envelope, and
the final collapse produces no outgoing shock. For example, the time
between first shock break out (t = 0 in \Fig{various}) and the launch
of the final outgoing shock in Model TH102p25 is only 64 days, whereas
the ``duration'' in Table 3 is 80 days. On the other hand, light can
continue to diffuse out and shells continue to collide long after the
core has collapsed and that lengthens the duration. The ``M'' models
differ from the ``TH'' models in having a higher surface boundary
pressure and greater mixing length parameter. The resulting small
radius leads to a light curve that is initially fainter, but not
qualitatively different after the first two weeks.

The duration and luminosity of some of the models in \Fig{various},
especially TH102p25 and M108p85, are in reasonable agreement with the
observations of SN 1961V. There is also time variability due to repeated
pulses, and a tendency of the luminosity to rise at late times as the
pulses themselves become more violent. The heavier models, TH102p25
and TH103p25, also show ``tails'' resulting from continuing shock
interaction after core collapse. It is important to recognize that no
$^{56}$Ni is ejected in any of these models; all of the observed
luminosity is from diffusion out of shock-heated ejecta or CSM
interaction. Still heavier models seem to be active too long, though
see \Sect{long} and \Sect{multiple}.

The shape of the light curve after the final brightest peak on the
plateau is not well calculated in some of the models, especially the
narrow spikes at 250 days in Model TH103p25 and 300 days in Model
TH102C, and the ledges at the end of the plateau just before abrupt
plunges by an order of magnitude in luminosity. The narrow spikes are
artifacts of a crude treatment of radiation transport and shock
hydrodynamics in a 1D model. As we shall see repeatedly, when the
shock snowplows in a region of decreasing $\rho r^3$, with $\rho$ the
density and $r$ the radius, matter piles up in a geometrically thin
shell, all moving at the same speed. In a multidimensional calculation
this pile up is not nearly so extreme \citep{Che82,Che16} and the
medium fragments into a clumpy shell with a thickness more comparable
to 10 - 20\% of its radius. As it is in 1D, when two very thin shells
collide, the inelastic collision results in a discontinuous burst of
energy as all their differential kinetic energy is abruptly converted
to light. Brief spikes like the one in the late time light curve of
TH103p25 should thus really be much broader and fainter. Thus, the
absence of such narrow luminosity spikes in the observed light curve
is not really in contradiction with these models. Similarly, the very
sharp ledge of late time luminosity might be a numerical artifact. It
comes from energy that has been radiated into the interior by a shock
diffusing out again. The inward and outward transport of this
radiation is not handled well in a Lagrangian code with a fixed inner
boundary when the interior of the star has, in fact, already collapsed
to a black hole. We suspect this late time plateau in luminosity is an
overestimate, but a more realistic calculation is needed. With these
adjustments, the agreement between observations and e.g., Model TH102p25
would improve at late times.

\begin{figure}
\includegraphics[width=\columnwidth]{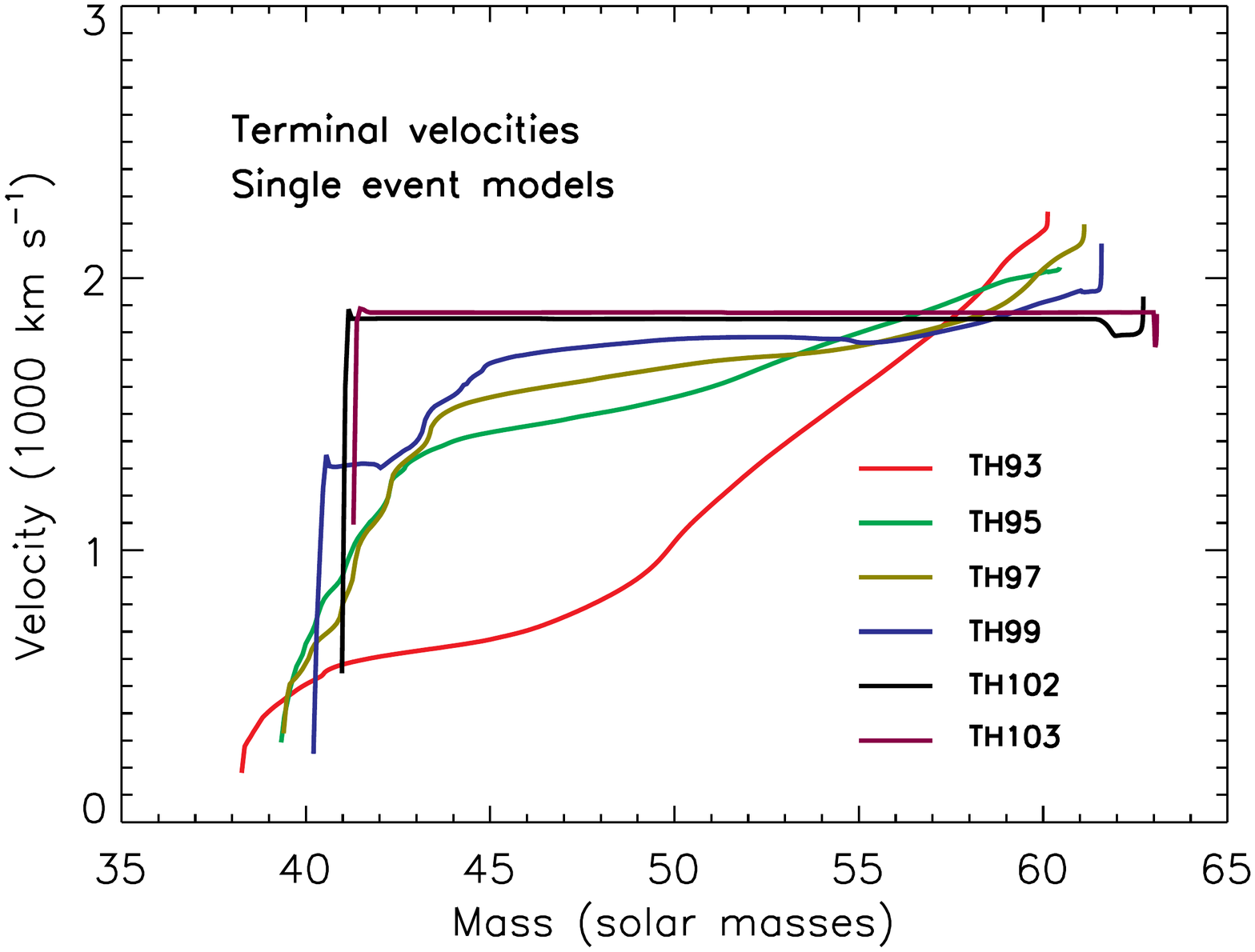}
\includegraphics[width=\columnwidth]{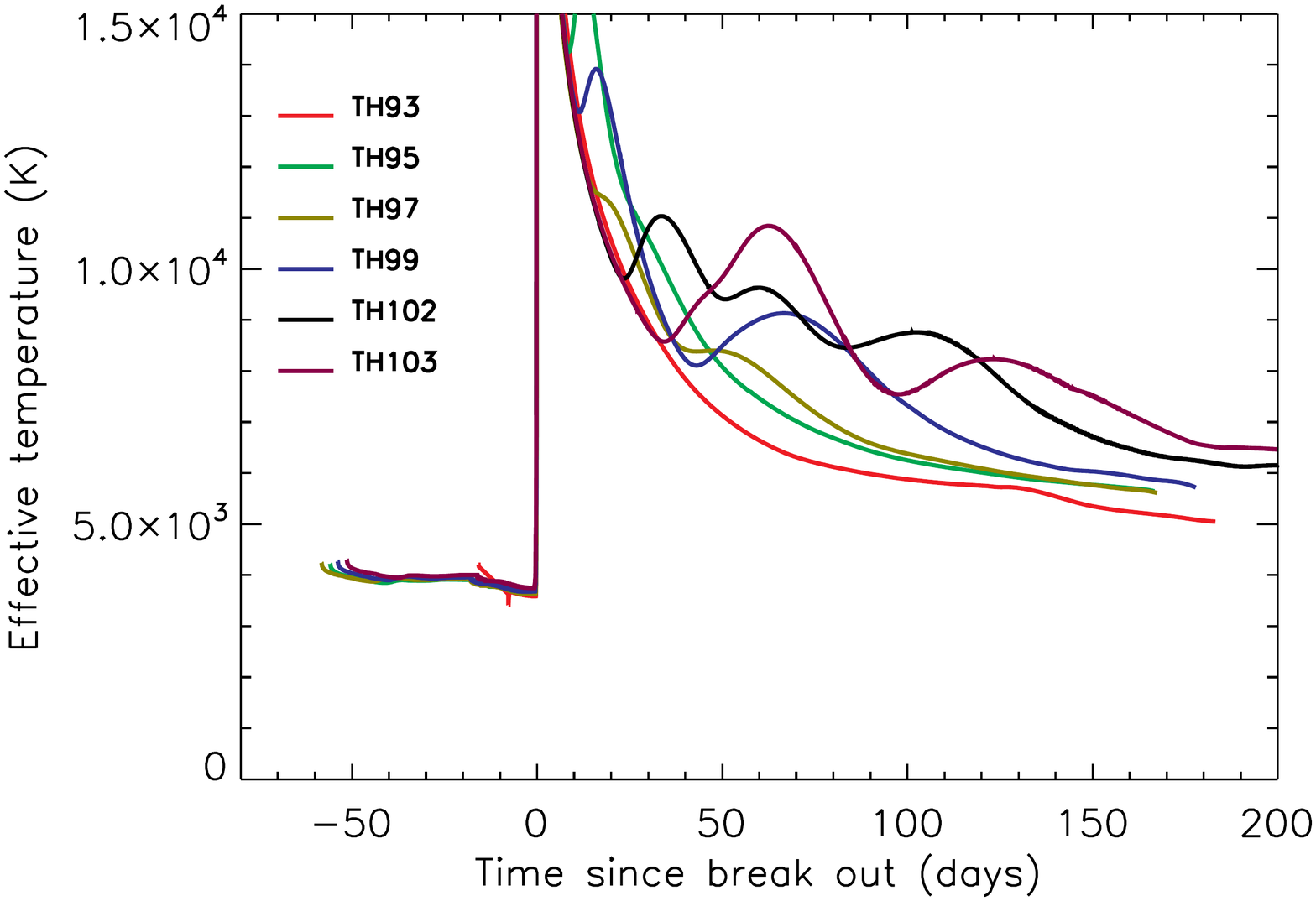}
\caption{(top:) Terminal velocities and (bottom:) effective emission
  temperature histories of Models TH93p25, TH95p25, TH97p25, TH99p25,
  TH102C and TH103p25 during a time when the electron-scattering
  photosphere is well defined. Lower mass explosions resemble typical
  Type II-P SNe, albeit with a low energy to mass ratio. The
  higher mass models show distinctive signatures of pulsationally
  generated shock heating. At late times all the ejected matter ends
  up coasting at the same speed. This is an probably an artifact of
  the 1D calculation. During a pulse, the luminosity increases at
  nearly constant radius so the temperature rises. \lFig{teff}}
\end{figure}

On closer inspection though, the ``TH'' and ``M'' models have
problems. The ones with sufficient duration and velocity are just too
bright, regardless of initial radius. The narrow width of the brightest
observed peak is not reproduced, even if the questionable ledges of
late time emission in the models are removed. The brightening a year
before SN 1961V and multiple years of ``afterglow'' are not explained
without invoking additional processes.

Better agreement can be achieved using the lower energy, high-carbon,
A-series models, and a smaller radius for the pre-SN
star. \Fig{singlepeak} shows the pulsing history and bolometric light
curve of one such model, A123p65. This model has fewer, more widely
spaced pulses than TH102p25, and the light curve thus shows less rapid
variability. The smaller radius, by a factor of two, makes the initial
display fainter, but most importantly, the larger carbon abundance
weakens the pair instability. Overall, the kinetic energy of the
explosion is reduced by almost a factor of two (\Tab{tmodels}) and,
along with it, the total amount of radiant energy emitted by the SN.
This brings the model light curves into much better agreement with
SN~1961V.  Model TH102p25 radiates $1.26 \times 10^{50}$ erg while
Model A123p65 radiates only $7.3 \times 10^{49}$ erg, though this is
still substantially more than observed for SN 1961V (\Tab{61vlite}) if
no bolometric correction is assumed. The necessary helium core mass
for instability is also raised from 46.14 \Msun \ to 55.15 \Msun \ and
the main sequence mass goes from 102 \Msun \ to 123 \Msun. The black
hole remnant mass increases from 40.8 \Msun \ to 51.2 \Msun.

\Fig{singlepeak} shows the velocity and local luminosity in Model
A123p65 at several relevant times in its evolution. During the early
plateau stage, not shown in \Fig{singlepeak}, the photosphere lies in
the outer ejecta where the speed is near 2000 km s$^{-1}$. During the
brief period between core collapse and the beginning of the last major
peak, the photosphere briefly recedes to lower velocity, although
spectral lines would still be affected by the faster moving material
farther out. With time, and especially near peak, the shock moves into
higher velocity ejecta and eventually overtakes the
photosphere. Brief spikes in the light curve at 140 d and 200 d
post-peak luminosity are due to colliding thin shells, the sharpness
of which is overestimated in the 1D study.

The final velocities, after all shock interaction is over, are close to
2000 km s$^{-1}$ (\Fig{teff} and \Fig{singleun}).  TH102p25 had an
envelope mass of 16.6 \Msun; A123p65 had a lower-mass H envelope, 10.0
\Msun, but less kinetic energy. Testing the effect of a single
parameter is difficult in these models because so many attributes are
interconnected. For example, changing the envelope mass also changes
the dynamics of the explosion and the pulse history. Changing the mass
loss rate changes the envelope mass, but also the helium core mass.
Decreasing the envelope mass of A123p65 would assist with narrowing
the light curve at peak, but might also lead to longer gaps in the
emission at earlier times.

While Model A123p65 is a phenomenal fit to the light curve, it has
some lingering issues.  The model had velocity much greater than 2000
km s$^{-1}$ in its interior close to the time of peak luminosity
(\Fig{singleun}), but this high speed was buried beneath optically
thick material making Zwicky's observation of 3700 km s$^{-1}$ somewhat
problematic.  Both the brightening 366 days before peak and the
subsequent years of enduring emission near the Eddington luminosity
are unexplained. The pre-SN point on date 1960.942 might
indicate activity as a LBV, as suggested by \citet{Goo89}, or some
other phenomenon associated with the rapid Kelvin-Helmholtz
contraction of the star following central carbon ignition and prior to
oxygen burning.  To be fair though, these issues would be vexing for
any other explosion mechanism as well, and invoking such additional
precursor variability and late-time CSM interaction is well justified,
even if it is not uniquely predicted by the PPISN models.

It is certain, for example, that the very luminous progenitor star
was rapidly shedding mass when it died. Reducing the hydrogen envelope
to $\sim$10 \Msun \ during the 300,000 years that helium burned
requires an average mass-loss rate greater than 10$^{-4}$ \Msun
\ y$^{-1}$. This is easily achievable for the normal winds of luminous
LBV-like stars \citep{SmiOw06}, even without their eruptive mass
ejection.  Prodigious mass loss may also be assisted by close binary
interaction, as is assumed to be the case for the precursor
variability and mass loss before the eruption of Eta Carinae
\citep{Smi11b,Smi18}. Given the necessity of some sort of surface
activity to explain the brightening a year or more before explosion,
and the superheating of the hydrogen burning shell by core
contraction, the actual mass-loss rate was probably much larger near
the end. Taking a representative value of 0.001 \Msun y$^{-1}$ and a
wind speed of $v_{\rm wind}$ = 100 km s$^{-1}$, plus a velocity for
the outer few hundredths of a solar mass of ejecta of $v_{\rm shock}$
= 2500 km s$^{-1}$, as the models find, implies a CSM interaction
luminosity of $L_{\rm CSM} \approx 0.5 \dot M v_{\rm shock}^3/v_{\rm
  wind} \sim 5 \times 10^{40}$ erg s$^{-1}$, lasting for a year or
so. The luminosity would gradually decline as the fastest moving
ejecta were decelerated, or when the edge of the CSM shell was
reached.

Altogether, we find that, if it is permissible to invoke precursor
LBV-like variability to explain progenitor variations before SN~1961V
and late-time CSM interaction (with dense CSM resulting from that same
precursor variability) to explain the very late-time luminosity, then
a PPISN model can do an excellent job of matching the very unusual
main light curve, low observed expansion speeds, and low kinetic
energy of SN~1961V.

\begin{figure}
\includegraphics[width=3.1in]{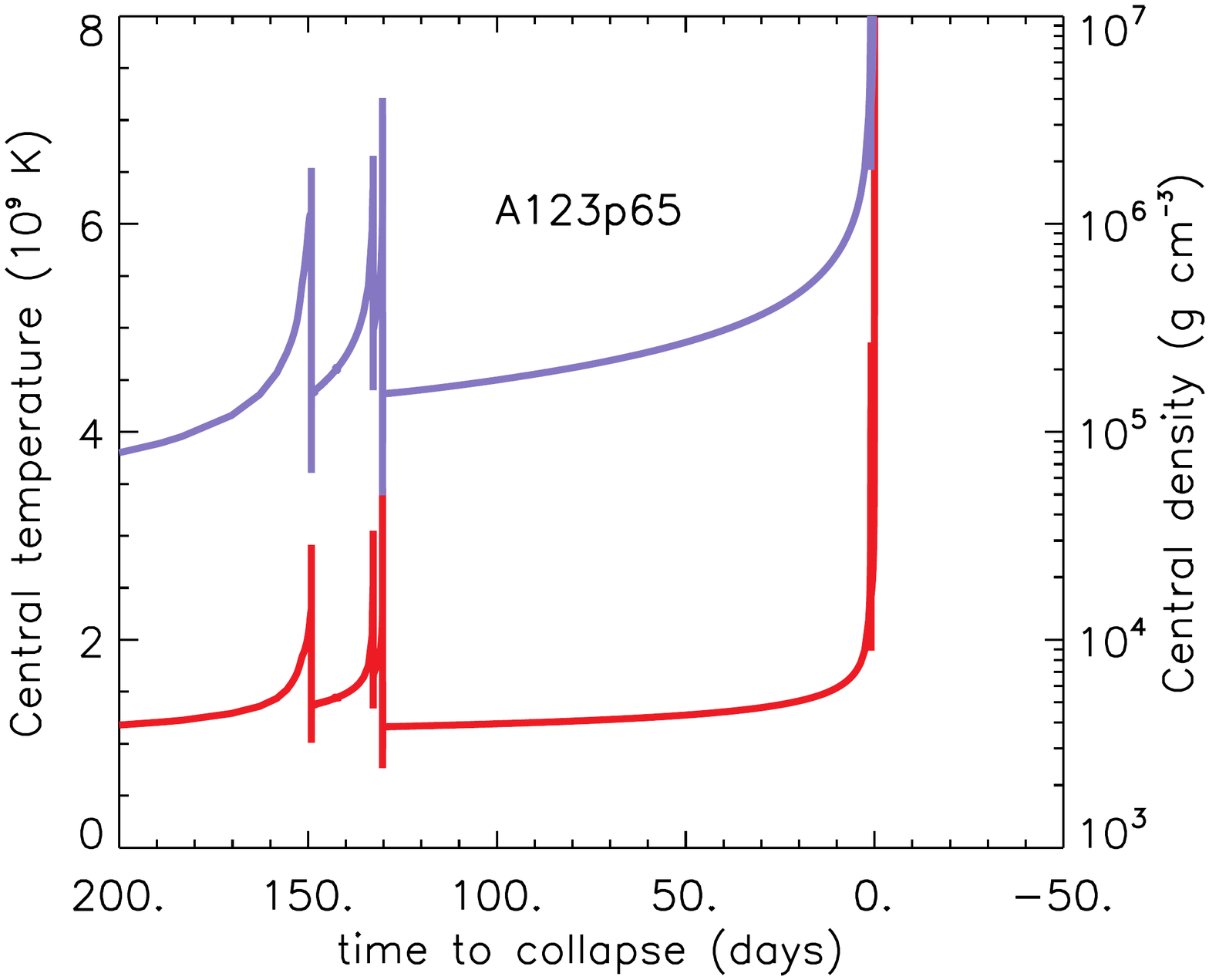}
\includegraphics[width=3.0in]{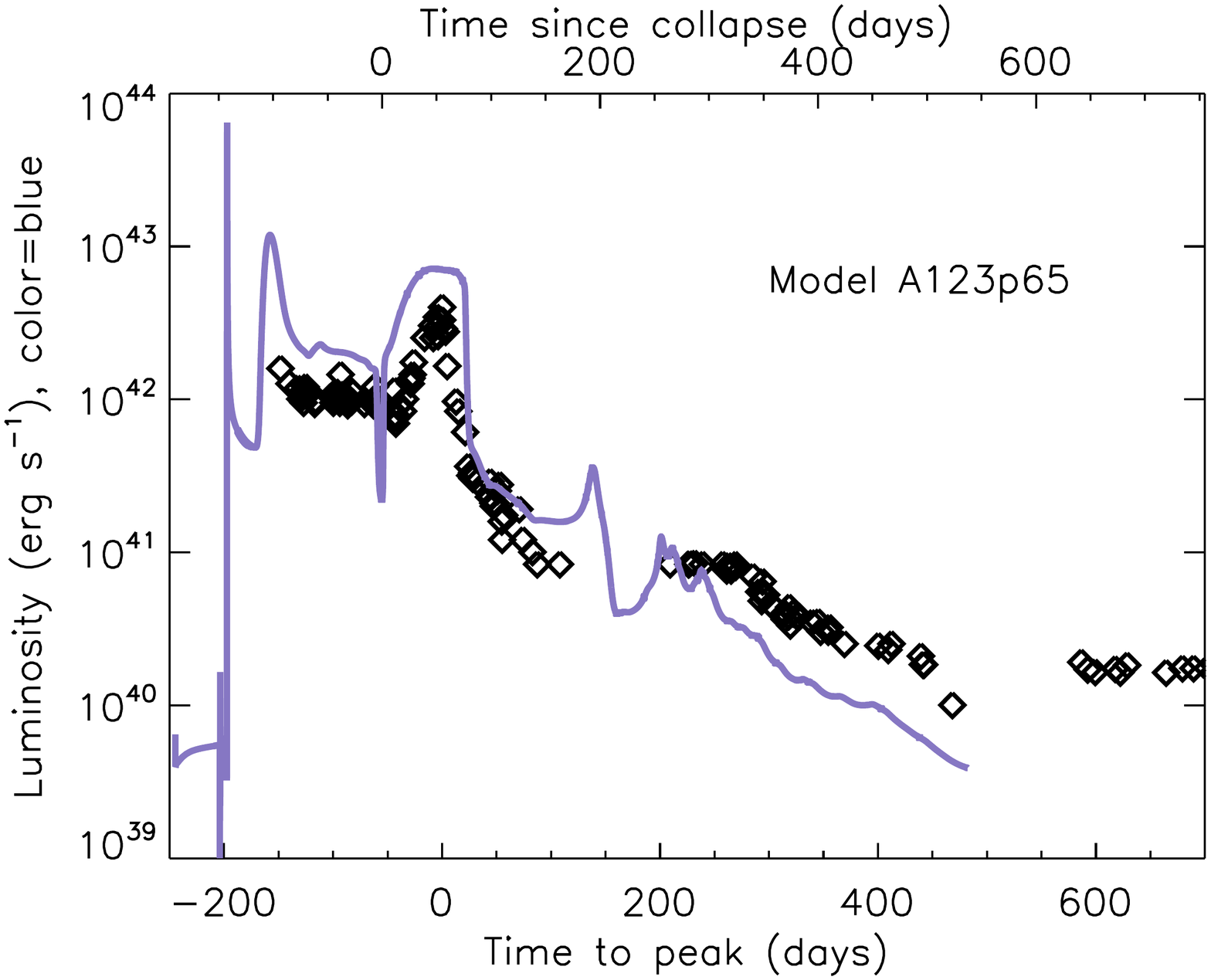}
\includegraphics[width=3.1in]{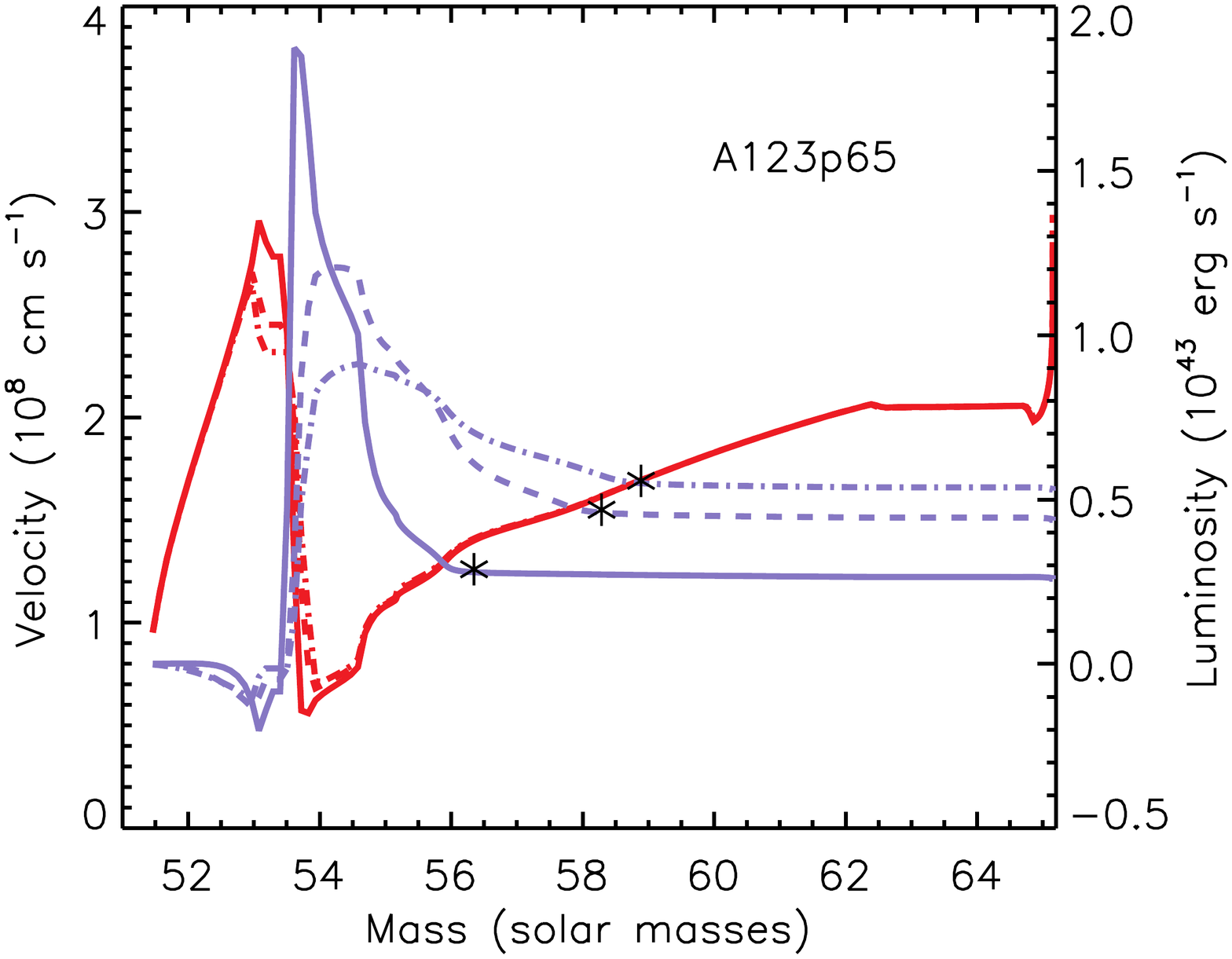}
\caption{(top:) Pulsing history of Model A123p65. The central
  temperature and density are shown as a function of time to core
  collapse. A weak pulse ($T_{p} = 2.91$ GK) at 149.1 days before
  collapse launches a shock that unbinds most of the envelope. Two
  stronger closely separated pulses at 132.8 days ($T_{p} = 3.05$ GK)
  and 130.3 days ($T_{p} = 3.46$ GK) illuminate the previously ejected
  material.  Finally a series of three rapid pulses 0.977, 0.958 and
  0.925 days ($T_{p} = 4.04$, 4.25 and 4.86 GK) leads to a terminal
  brightening just as the core collapses to a black hole (the
  brightening is delayed from the time of pulses because of diffusion
  time). (middle:) Comparison of the bolometric light curve of Model
  A123p65 with that estimated for SN 1961V.  Time is plotted relative
  to the observed peak for SN 1961V on the bottom axis and to core
  collapse on the top axis. The single observed point at -366 days is
  omitted in this plot as are observations beyond 700 days
  postpeak. (bottom:) Radiation transport in Model A123p65. The
  velocity (red) and luminosity (blue) are shown as a function of mass
  for three different times, 10 d (solid), 20d (dashed), and 25 d
  (dash-dotted) after core collapse (see the time scale on the top of
  the middle panel). Note the diffusion of light ahead of the
  shock. Asterisks indicate location of the electron scattering
  photosphere.  \lFig{singlepeak}}
\end{figure}

\begin{figure}
\includegraphics[width=\columnwidth]{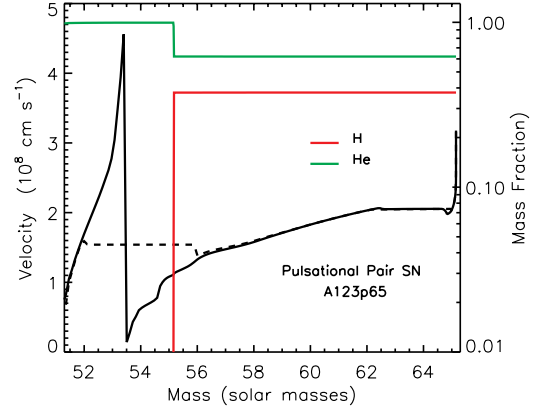}
\caption{Velocity (black line) and ejected H/He composition for
  Model A123p65 as a function of mass. The velocity at the time of
  iron-core collapse (solid line) and at 700 days (dashed line) are
  shown. The photosphere is just outside of the high velocity material
  in the pre-SN star. Velocities in the hydrogen-rich material
  range from 1500 to 2100 km s$^{-1}$, except for a small amount of
  high velocity material ($\sim$3,000 km s$^{-1}$) at the outer
  edge. The plot for Model A123p7 would be similar, but the final
  velocities are about 15\% greater. \lFig{singleun}}
\end{figure}

\subsection{Repeating Widely Separated Events}
\lSect{long}

A unique characteristic of some PPISNe is their ability to produce
violent explosions separated by decades, or even millennia of
quiescent, star-like behavior \citep{Woo17}. Much has been made of the
possibility that the same star that produced SN 1961V might still be
``alive'' decades later or even now. Occasionally, that possibility
has been used to argue against a SN origin for the main event
\citep{Van12}. Conversely, the presence of a strong radio or x-ray
source has been taken as evidence that a real SN happened
\citep{Chu04}, and that the star must therefore be gone. Both
arguments are potentially specious if a star can explode violently as
a SN more than once.

\Fig{doublepeak} shows two examples. Both are of the low carbon
variety, TH114p25 and TH115p25. It is possible to find examples among
the high-carbon models, e.g, A140p6, that have a similar history and
initial light curves comparable in luminosity and duration to SN
1961V, but, in this limited survey, none was found with more than one
pulse in the initial outburst. Single-pulse events end up looking like
ordinary PISNe (\Fig{ppilite}) and were
excluded.

The light curves for TH114p25 and TH115p25 in \Fig{doublepeak} are
really just the first outbursts of these models. Model TH114p25 was a
SN again and finally collapsed to a black hole 40 years later (i.e.,
in 2001). Model TH115p25 did the same thing 106 years later (i.e., in
2067).  No great weight should be given to these specific dates, but
the implication is that the progenitor of SN 1961V might or might not
still be there and that it could still die at anytime. The second
events are probably not very bright optically. They eject no hydrogen
or radioactivity, and the hydrogen shells from 1961V itself are
now out at $4 \times 10^{17}$ cm, so any interaction is likely to be
long, faint, and not optically efficient.

\begin{figure}
\includegraphics[width=\columnwidth]{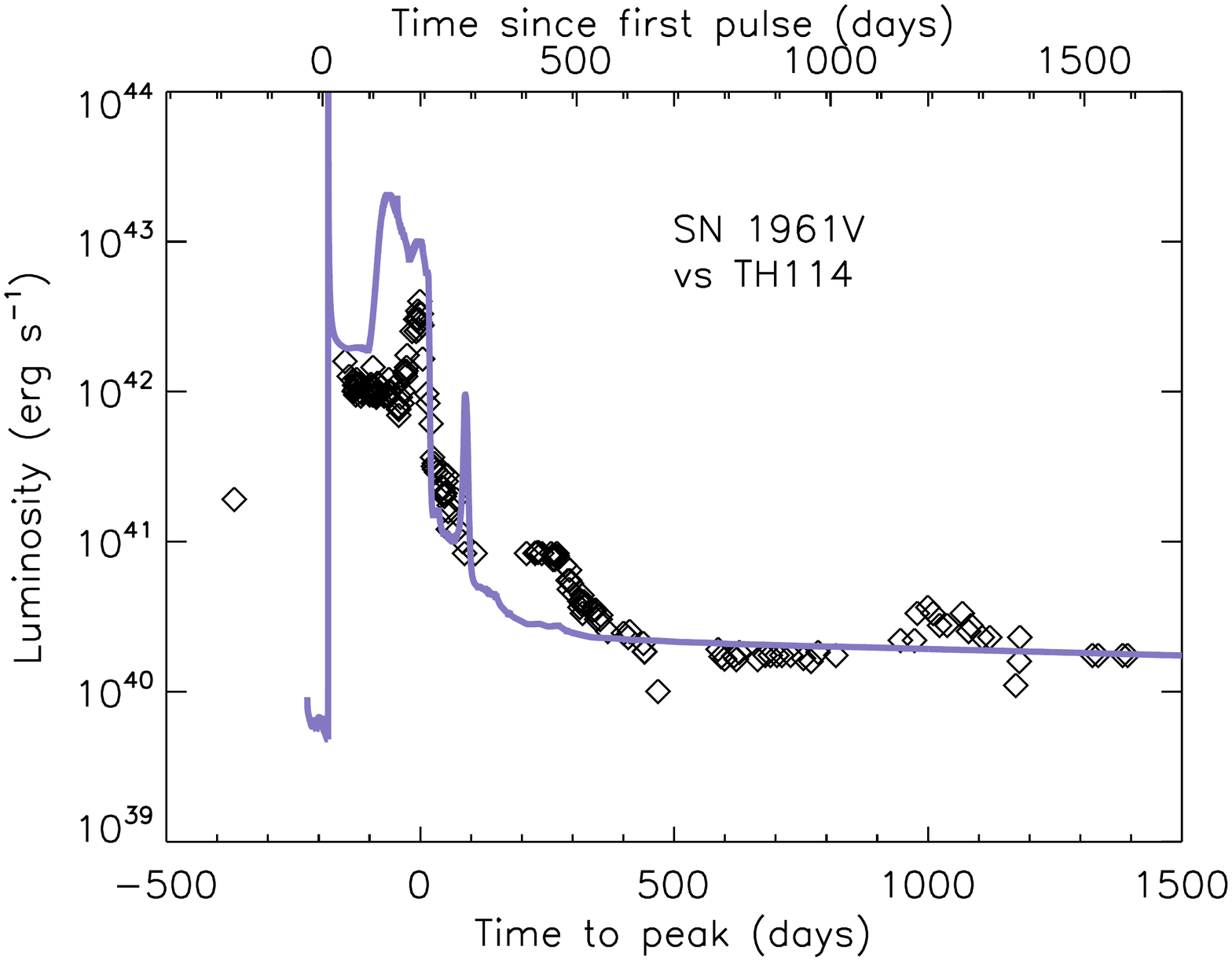}
\includegraphics[width=\columnwidth]{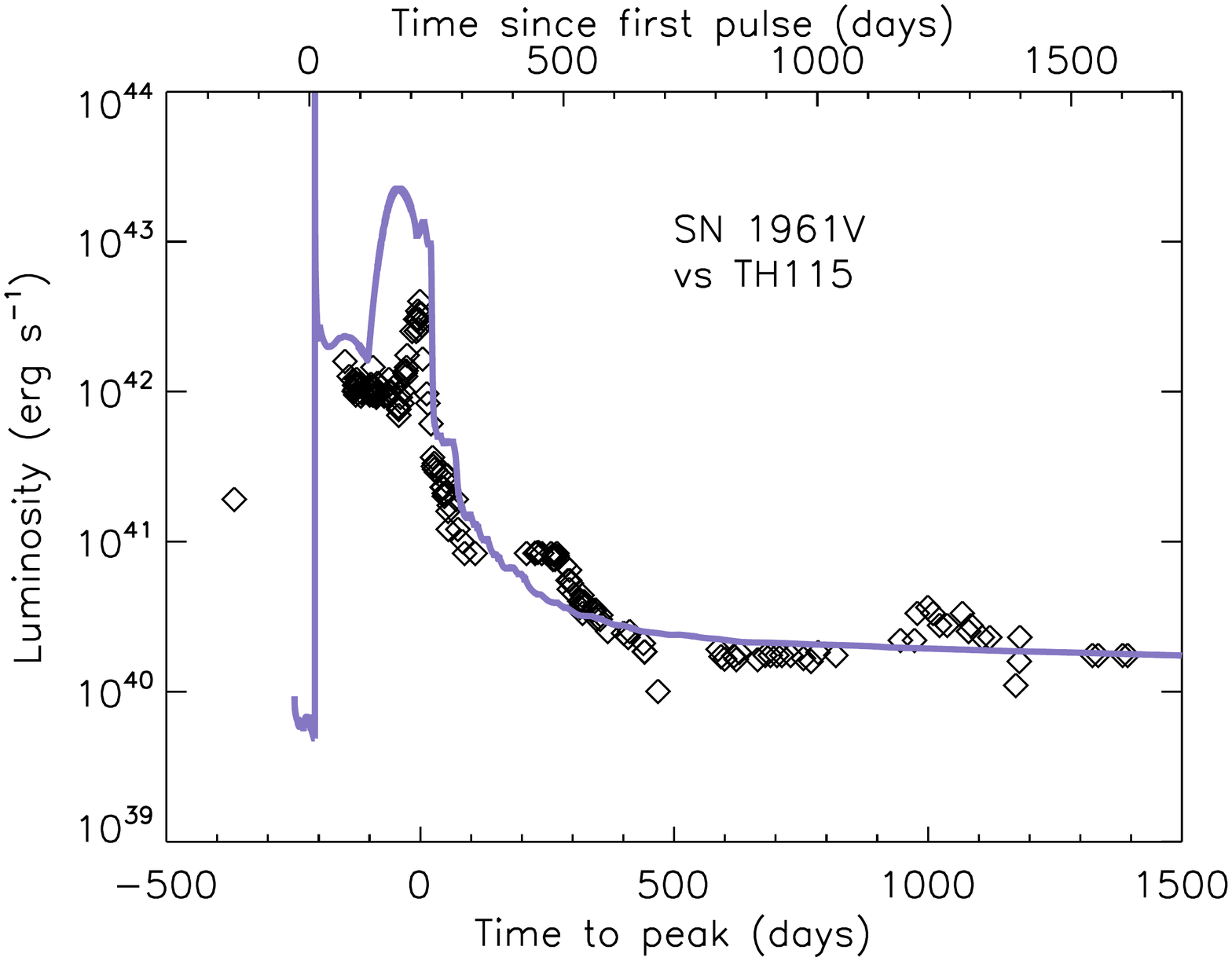}
\caption{(top:) The early bolometric light curves of Models TH114p25
  and TH115p25 compared with SN 1961V for the 4.5 years when the SN
  was well observed. Time for the model is measured since the first
  pulse (top axis) and for the SN relative to the time of its observed
  peak luminosity (bottom axis). The model luminosity prior to t = 0
  is the pre-SN star. Emission after about 500 days is from the
  residual stars, both of which retain a mass at this point of 50
  \Msun.  The overall light curve of SN 1961V is in good agreement
  with this model, though not the initial observed point at $-$366
  days, which must have other causes.  In Model TH115p25, a stellar
  remnant would still be shining at approximately the Eddington
  luminosity ($\sim10^{40}$ erg s$^{-1}$) today, but for Model
  TH114p25 the star died around 2001.\lFig{doublepeak}}
\end{figure}

\begin{figure}
\includegraphics[width=\columnwidth]{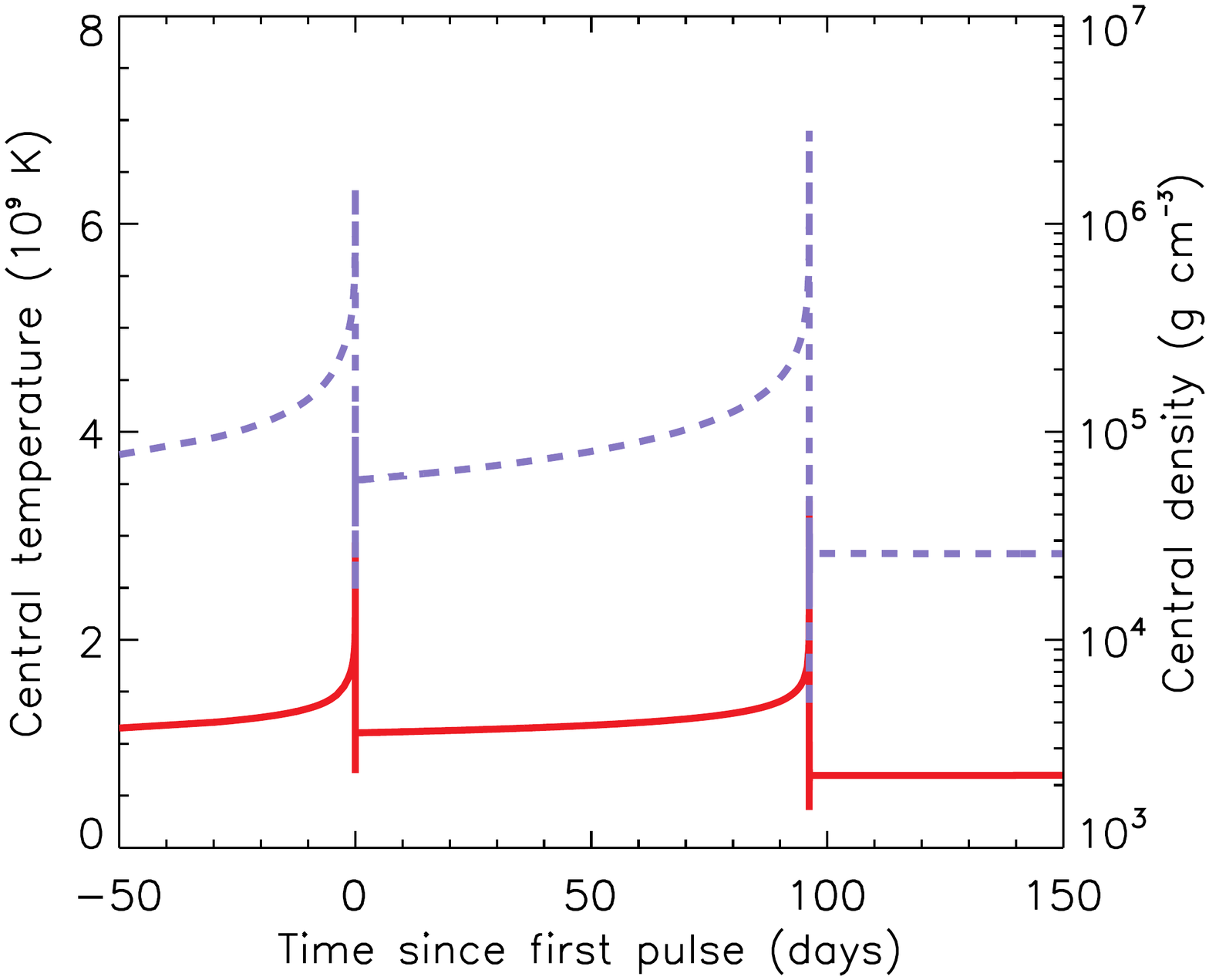}
\includegraphics[width=\columnwidth]{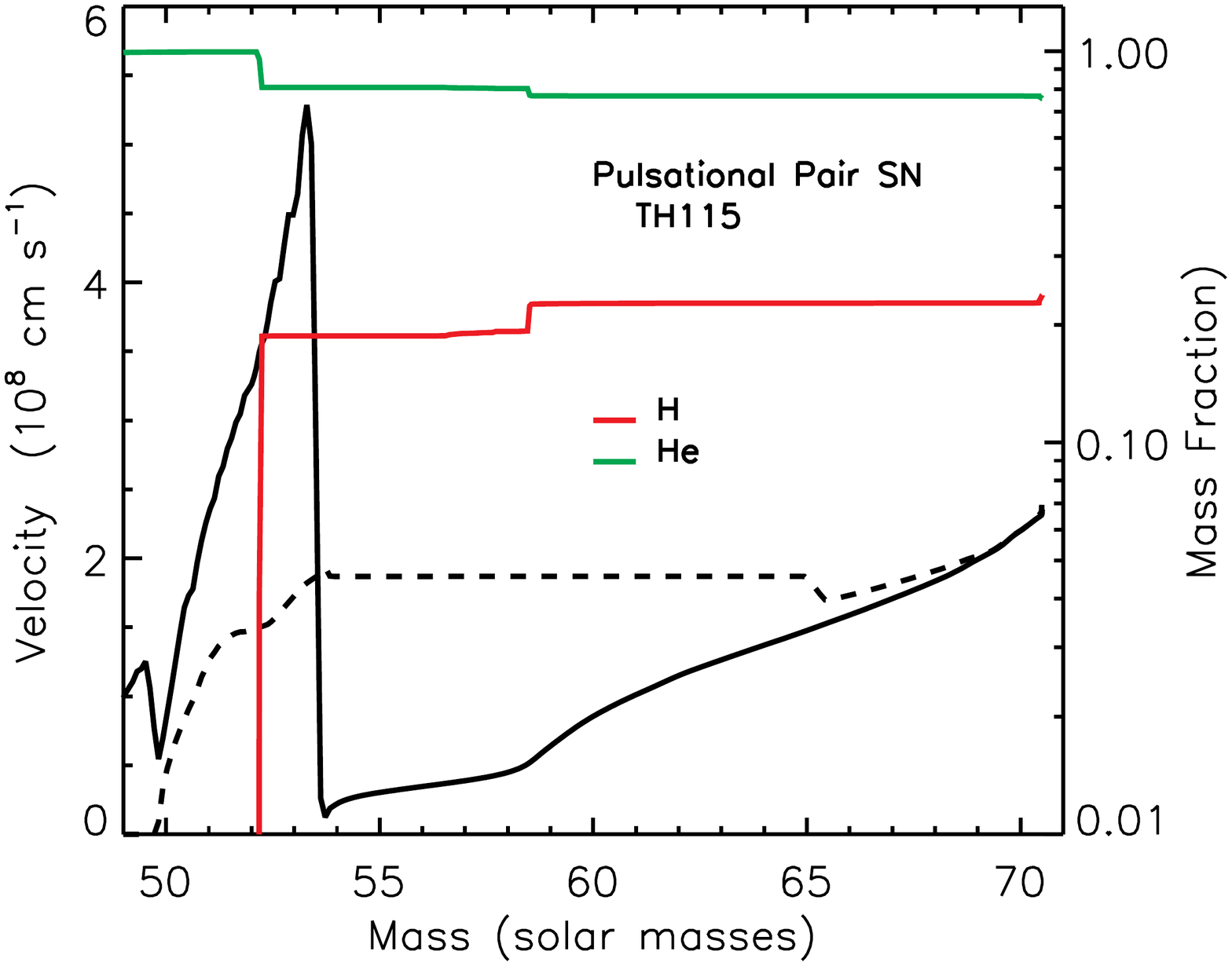}
\caption{(top:) Pulsing history of Model TH115p25 during the first
  series of pulses. Additional events occurred 106 years later,
  culminating in the collapse of the core to a black hole. Central
  temperature and density are plotted vs time since the first
  pulse. The first pulse ejected the hydrogen envelope, producing a
  low-energy SN~II-P. The second pulse ejected matter that interacted
  with the prior ejecta to produce the late peak in the light
  curve. (bottom:) The velocity and composition profile for Model
  TH115p25 at a time just before peak (solid line) and much later when
  the luminosity has declined to 10$^{41}$ erg s$^{-1}$. The high
  velocities at the earlier time are well beneath the photosphere.
  \lFig{115tnun}}
\end{figure}

The speeds of the hydrogen-rich ejecta in these models are again close
to 2000 km s$^{-1}$ after 1961 (\Fig{115tnun}), and each eventually
makes a black hole of 46 \Msun.  These models have the additional
merit over those of \Sect{single} or \Sect{multiple} of explaining the
years of post-1961 emission of SN 1961V. The remaining star was
greatly inflated by the initial explosion. The central density and
temperature of the 49.9 \Msun \ remnant of Model TH115p25 after its
first couple of pulses are $2.6 \times 10^4$ g cm$^{-3}$ and 0.697
GK. The long lifetime of the stellar remnant is because this is near
the temperature where neutrino losses become less efficient than
radiative losses. The photosphere of the remaining star is almost
entirely helium. Indeed the hydrogen-poor photosphere of the remaining
star is a characteristic of any SN origin for the main event.
\citet{Van12} claimed that 61V was ``alive'' based on a lingering
source (Object 7) that had H$\alpha$ emission like an LBV wind, but
any survivor of the 1961 explosion would not have a hydrogen envelope
anymore.

During the long wait before its final collapse, the star contracts,
with no nuclear burning, and radiates near the Eddington luminosity,
$\approx 1 \times 10^{40}$ erg s$^{-1}$. There may be small additional
contributions from (Eddington-limited) accretion from fallback and
circumstellar interaction with a wind, but a persistent luminosity
near 10$^{40}$ erg s$^{-1}$ seems guaranteed for many decades. The
emission properties of the central star are difficult to predict due
to fallback, the contraction of the still hot outer layers of the
protostar-like object, and the possible interaction of any
radiation with the SN ejecta, but it would be very hot with a radius
of order 10$^{12}$ cm and effective temperature $\sim50,000$
K. Without a hydrogen envelope, this luminous contracting He star
would resemble a very luminous Wolf-Rayet star, and may be faint in optical images
despite its high bolometric luminosity.  Recent limits on bolometric
luminosity, including any non-optical mission (especially deep UV
observations), substantially less than 10$^{40}$ erg would rule out
models like TH115p25 that still survive as a star today, but would not
constrain models that already died like TH114C.

\subsection{Enduring Explosions}
\lSect{multiple}

There is a third class of models in which violent pulsations persist
for several years, but not decades, and SN 1961V happened to be caught
in the later stages of the star's death. These models have very
irregular light curves, resulting from multiple colliding shells with various
speeds and masses, and obtaining a good match to SN 1961V is
difficult.  In all these cases, there would need to have been an
earlier bright transient, comparable to the brightness of the main
peak in late 1961, that occurred before the plateau phase of SN~1961V,
but was not observed.  Examples of this class are Models TH109p25 and
A126p65 (\Tab{tmodels}; \Fig{multiple} and \Fig{a126p65un}).  The core
in these models explodes several times. For TH109p25 there were three
pulses that happened 594 d, 358 d, and 250 d prior to core
collapse. The central temperature, after oscillations following the
first pulse damped, was 1.03 GK and after the second pulse, 1.15
GK. These temperatures set the waiting time for the next pulse to be
236 and 108 days, which is about the time for matter moving at 2000 km
s$^{-1}$ to coast to a few times 10$^{15}$ cm and become optically
thin. This gives rise to pronounced, multi-peak light curves that are
not as smooth as other models.  Much greater time or speed, and the
collisions would happen outside 10$^{16}$ cm where the optical
efficiency might be less and the collision times longer. Much less and
there would just be one continuous event like in \Sect{single}.
Similarly, in Model A126p65 there were also three major pulses
(\Fig{a126p65un}) prior to core collapse: a single weak pulse 381 days
before collapse (subsequent central temperature 1.04 GK); a double
pulse at 33.2 d and 30.1 d (subsequent central temperature 1.42 GK)
and a final pulse 0.85 days before core collapse.

The broad peaks in both Models TH109p25 and A122p65 before core
collapse are, as usual, due to individual pulses charging overlying
optically thick matter with radiation that then diffuses out, but the
narrow peaks after -150 days (relative to peak) result from colliding
shells at lower optical depth. The narrow 30 day peak of SN 1961V in
both models, results from ``circumstellar interaction'', not diffusion
and recombination or radioactivity. Its spectrum might thus be
novel. As remarked previously (\Sect{single}), these narrow spikes
in emission should actually be fainter and broader because of
multidimensional instabilities not followed in KEPLER.

\begin{figure}
\includegraphics[width=\columnwidth]{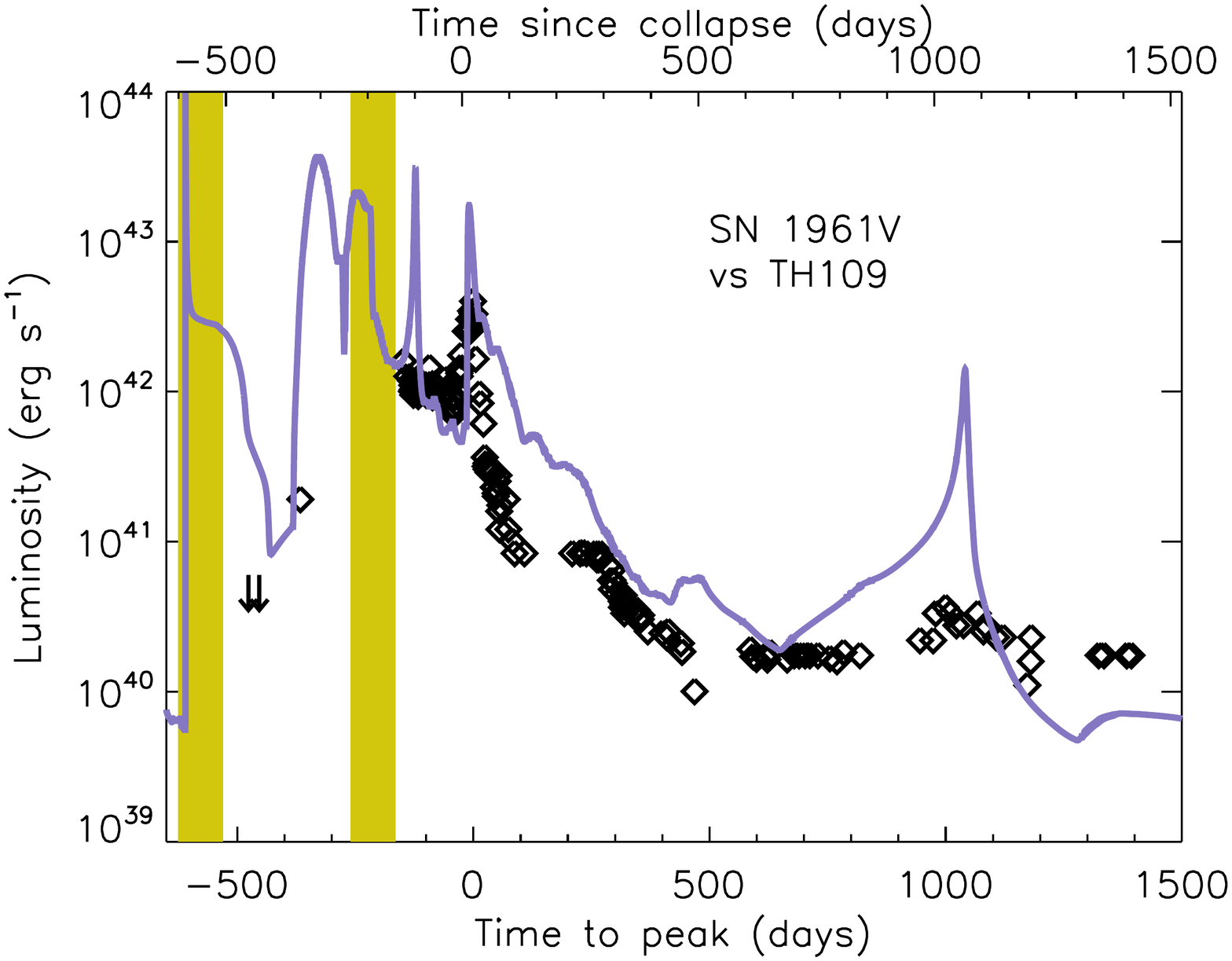}
\includegraphics[width=\columnwidth]{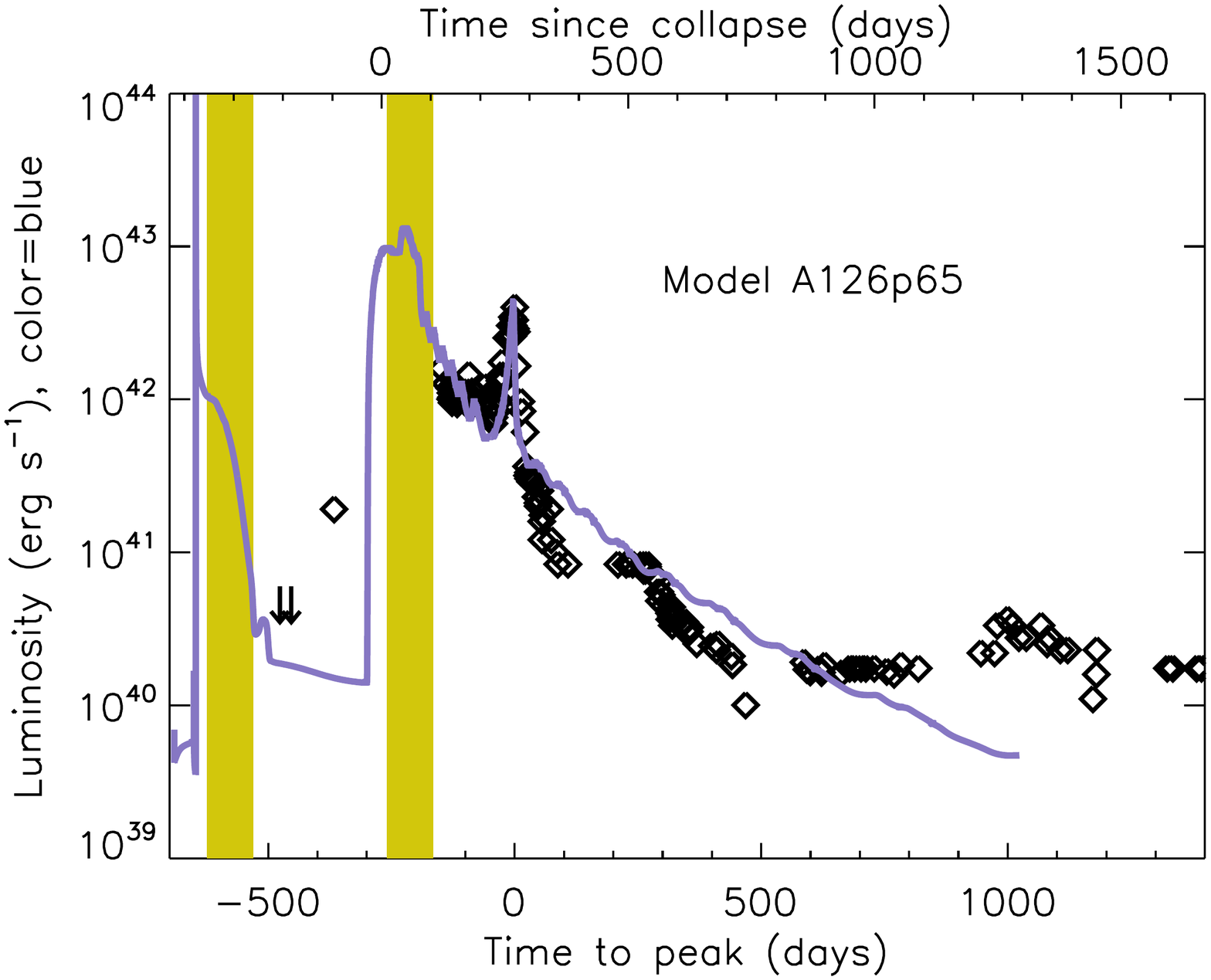}
\caption{Comparison of the bolometric light curves of Models TH109p25
  and A126p65 with observations of SN 1961V.  The model luminosities
  have a complex structure resulting from both repeated explosions and
  shell collisions. The models predict bright prior events that might
  have escaped detection. Note that days $-$260 to $-$165 and $-$625
  to $-$530 relative to peak (vertical gold bands) are when SN~1961V
  is not observable due to R.A. constraints (\Sect{absence}). Upper
  bounds on the luminosity at -453 and -476 days before peak can be
  problematic depending on the spectrum and exact timing of pulses. At
  time zero on the top axis the core collapses to a black hole and is
  removed from the calculation. Very late time emission might be
  augmented by interaction with a pre-SN stellar wind.\lFig{multiple}}
\end{figure}

\begin{figure}
\includegraphics[width=\columnwidth]{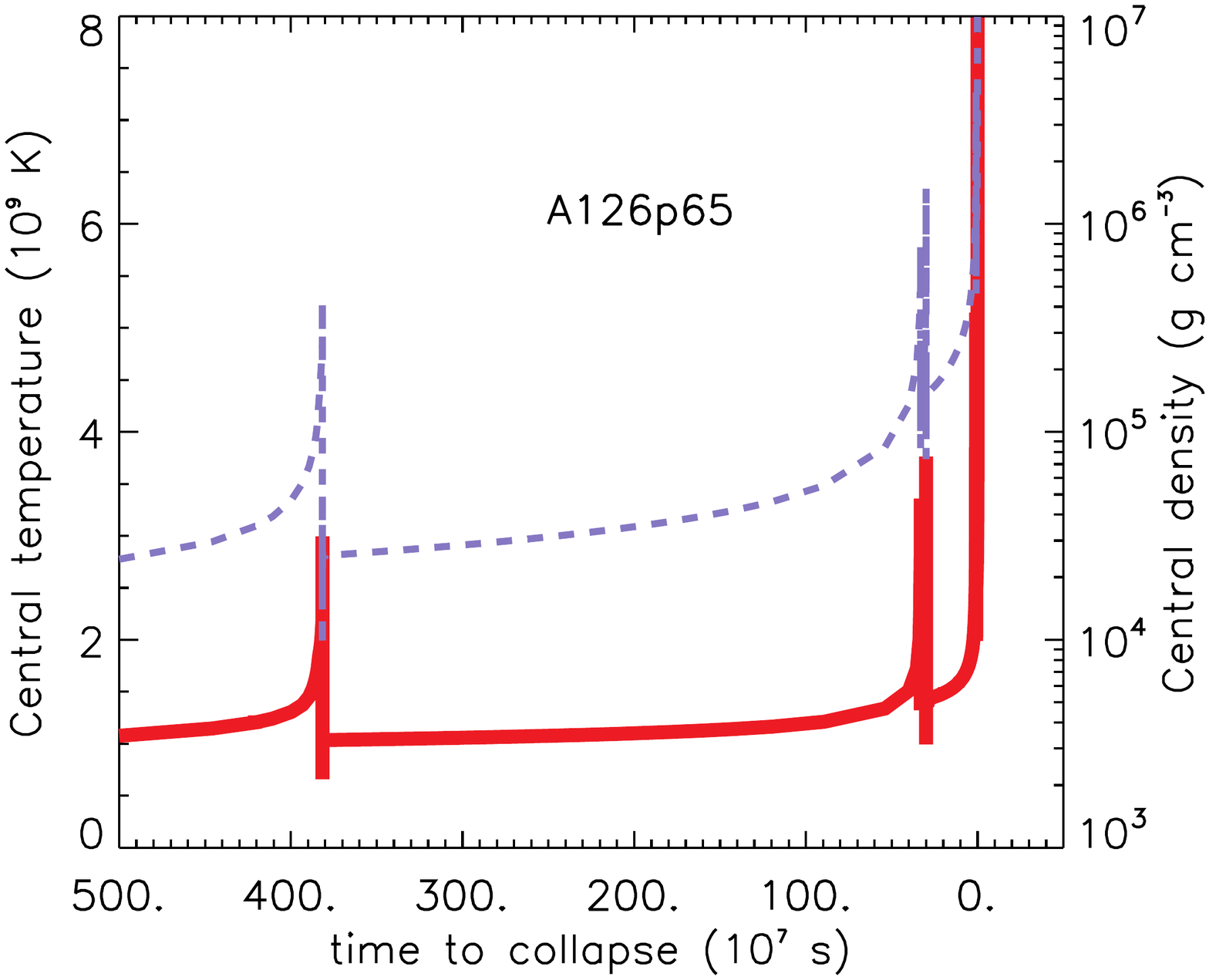}
\includegraphics[width=\columnwidth]{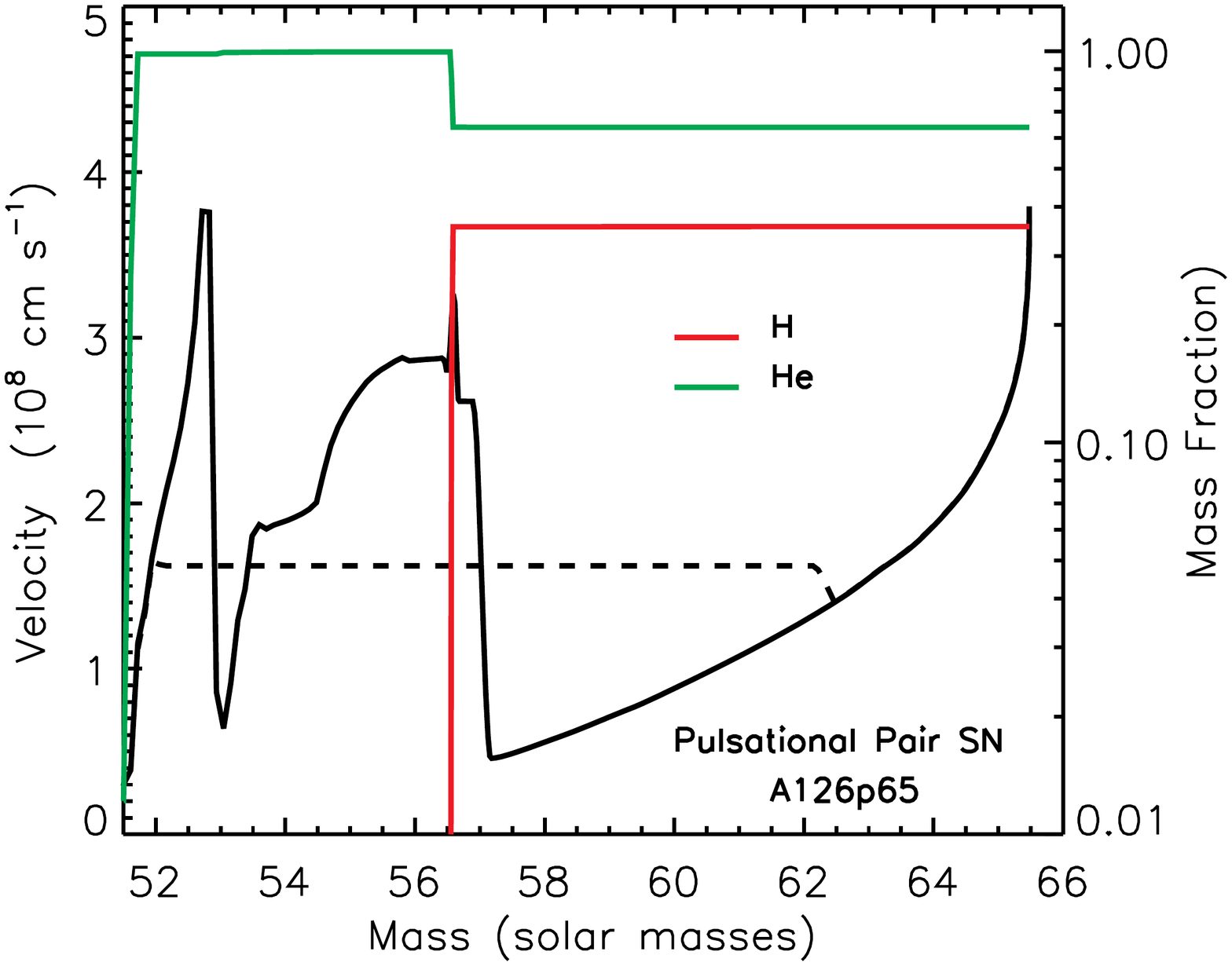}
\caption{(top:) The pulse history for A123p65 shows three pulses plus
  a final collapse to a black hole. (bottom:) ejected composition and
  velocity for the same model. The solid line shows the velocity at
  the time of core collapse with high values due to the terminal
  pulses shown in the top plot. The dashed line shows the velocity
  years later at the last point plotted in \Fig{multiple}.
  \lFig{a126p65un}}
\end{figure}

Still, the timing and qualitative agreement, especially of Model
A126p65, with SN 1961V for over 500 days of emission is
impressive. The late time emission comes from continuing shock
interaction, and could, of course, be enhanced further if there was
large mass loss due to precursor LBV-like variability, as noted
earlier. In Model TH109p25, there is an additional collision of shells
about 1000 d after peak. Once again this peak should be broader and
fainter because of multi-dimensional effects. Crudely, one might
expect the thin shell to have been spread over a region with thickness
$\Delta r/r \sim$ 10 - 20\% \citep{Che16} and the peak in the light
curve broadened by a comparable amount, $\Delta t/t \sim 10$ - 20\%
$\sim 150$ d.  Repeated outbursts on the tail of up to 1 mag were
reported by \citep{Ber70}.

Both these models also offer the possibility of explaining the
brightening one year before peak. Model TH109p25 accidentally passes
right through that point. In Model A126p65, an explosion had already
occurred, but the bolometric light curve had already fallen to an order
of magnitude fainter than the pre-SN limit. CSM interaction of the
leading shock with the pre-SN wind might explain the
discrepancy.

In any case, CSM interaction would need to be invoked in both these
models to explain the very late-time emission after 1000 days. An
increasingly uncertain bolometric correction must also be applied at
these late times, so the actual luminosity may need to be larger to
explain what was seen in optical wavelengths.

\subsection{Absence of Evidence for Pre-supernova Supernovae}
\lSect{absence}

Some of the model light curves that most closely resembled SN~1961V
also had ``pre-SN'' emission as bright as the main event in SN~1961V
that were, unfortunately, not observed.  In some cases these bright
peaks came shortly before discovery of SN~1961V at a time when data
was missing or sparse.  Remarkable examples are the bright initial
peaks seen in the ``single-event'' model A123p65 discussed in
\Sect{single} (Fig.\ 6), as well as the ``enduring explosion'' models
TH109 and A126p65 discussed in \Sect{multiple} (Fig.\ 10).

Without a more detailed examination of archival plates or observing
logs, and without entering the minds of observers at the time, it is
difficult to ascertain the implications of the lack of data.  At the
present time, with multiple ongoing professional and amateur SN
searches scouring the sky, it would be difficult to believe that such
a bright event would be missed.  For a bright event in 1960, however,
things could have been different.

In this regard, it is worthwhile to examine the observability of
SN~1961V.  For its location at R.A. $\approx$ 2.4$h$ and DEC $\approx
+37^{\circ}$, SN~1961V was close to the Sun and at high airmass ($>$2)
for northern hemisphere observatories from late March until
early July.  These dates translate roughly to a time period from day
$-$260 until day $-$165 relative to the main peak in November 1961.
Indeed, SN~1961V was discovered on July 11, 1961 \citep{Wil61},
just after it had emerged from behind the Sun, and the last
pre-SN observation was long before day $-$260 at day $-$366 relative
to the peak.  Therefore, if a very bright early peak occurred shortly
before discovery (as in models TH109 and A126p65 in Fig.~10), it would
have been very difficult or impossible to observe.  For this reason,
we conclude that the PPISN model light curves shown in Figures 6 and
10, which are predicted to have bright but unobserved early peaks
during this time period, provide viable explanations for SN~1961V.

Model A126p65 also shows another previous bright peak almost 2 years
before the November 1961 peak. During that time frame, the previous
period when SN1961V was unobservable was at day $-$625 to day $-$530
relative to peak; it is perhaps bad luck that this almost exactly
coincides with the first luminosity peak of Model A126p65 in Figure
10, prohibiting it from being observed.

We also note two reported observations of the source at SN~1961V's
position on 1960.641 and 1960.704 (see Table 1), which we choose to
plot as upper limits in Figure 10 at days $-$476.3 and $-$453.3.
These points would seem to disfavor model TH109, as they are somewhat
below the predicted luminosity at that time.  We caution the reader,
however, not to interpret these upper limits too rigorously.  These
points come from a private communication from P.\ Wild, as recounted
by \citet{Ber63}, stating only that the source was ``about 17 mag to 18
mag" on those dates.  However, no information about the weather or
seeing is available, nor the wavelength response of the emulsion used,
and this may have been near the limiting magnitude of the 16-inch Bern
Observatory Schmidt telescope used by Wild. As such, SN~1961V's
progenitor may have been very difficult to distinguish from its
surrounding association.  We prefer to interpret these reports as
loosely indicating that SN~1961V was not extremely bright at that
time.

\section{Conclusions}
\lSect{conclude}

If SN 1961V was a star over 100 \Msun \ that exploded as a SN, the
most likely explosion mechanism was thermonuclear, brought on by pair
instability. The large iron core masses and shallow density gradients
at their edges are unfavorable for neutron star production in such
massive stars \citep[e..g.,][]{Rah22} . A collapsar-powered explosion
or a millisecond magnetar would require an unusually large amount of
rotation for a star that retained its hydrogen envelope, but had also
suffered extensive mass loss. While some combination of pair
instability and black hole accretion at a low level after the main
event is not ruled out, and might even be necessary to explain the
sustained emission, we have focused here on ``pure'' pair-instability
models.

An ordinary PISN, one where the entire star explodes in a single
pulse, could explain the approximate luminosity, duration, and
velocities observed in SN 1961V (\Sect{pisn} and \Fig{ppilite}), and
at least the star would reliably blow up. But PISN are too bright if they
last long enough, and too brief if they are faint enough
(\Fig{ppilite}). With total kinetic energies approaching 10$^{52}$ erg
(\Tab{pisntab}), these models also struggle to keep most of hydrogen
moving slower than 2100 km s$^{-1}$. PISN models place an interesting
limit on the progenitor of SN 1961V though. Even for nuclear reaction
rates optimized to require large masses, the progenitor was not more
than 160 \Msun \ on the main sequence; and had a helium core mass,
when it died, of less than 77 \Msun. The bolometric pre-SN
luminosity was therefore less than $1.4 \times 10^{40}$ erg
s$^{-1}$. No model in this paper that looked at all similar to SN
1961V had a greater main sequence mass or pre-SN
luminosity. When examined closely though, a PISN origin for SN 1961V
seems unlikely. Better agreement can be achieved using pulsating
models.

Three classes of PPISN models were explored. In the simplest case, SN
1961V was a single event that happened in 1961 (\Sect{single}).  Its
light curve was powered by repeated flashes in quick succession, but
the iron core collapsed before the brightest emission (i.e., before
the observed flare that peaked on day 1961.943 in Table 1) began. This
kind of model resembles those for ordinary SNe~II where the energy
from neutron star formation is all deposited centrally within a few
seconds, but it differs in that the central energy source repeats
discontinuously throughout the light curve. That energy can also be
significantly augmented, at late times, by colliding shells
(\Fig{various} and \Fig{singlepeak}). These variations imprint
themselves on the light curve.

Reasonably good fits to the bolometric light curve and velocity were
obtained for a variety of single-event models (\Fig{various} and
\Fig{teff}), but the best fits were for models with one-third solar
metallicity, high carbon abundance, relatively small radii, a helium
core mass of about 55 \Msun, and a hydrogen envelope of about 10 \Msun
\ - Model A123p65 (\Fig{singlepeak}, \Fig{singleun}, and
\Tab{tmodels}). Six hundred days of bolometric emission are replicated
to about a factor of two. Excess emission at late times (around 300
days post-peak) could naturally be attributed to interaction of a few
hundredths of a solar mass of ejecta with a speed of 2500 km s$^{-1}$
crashing into a pre-SN wind of only 0.001 \Msun \ y$^{-1}$, or
perhaps alternatively to black hole accretion. Emission one year prior
to peak must find another explanation, but the star was rapidly
adjusting its central structure then and may have been an LBV
(\Sect{physics}).

Other variations on the PPISN theme also give impressive matches to
observational data. In the second class of models, SN 1961V was one of
two widely separated SNe (\Fig{doublepeak} and \Fig{115tnun}),
although the second might not have been a bright optical event. More
work on the emission properties of shocks in dense media with a radius
of $\sim$10$^{17}$ is needed. Here the low-carbon models seem to have
an advantage in that explosive oxygen burning commences promptly and
can, with some searching in model space, set up conditions for a more
powerful second pulse $\sim$100 days later. After this, there ensues a
long wait while the star recovers in a protostar-like contraction
phase, experiencing Kelvin-Helmholtz evolution as it emits at about
the Eddington luminosity. This hot ``protostar'' may have disappeared
into a black hole after a few decades or could still be there. Its
luminosity, integrated across all wavelengths, would remain close to
10$^{40}$ erg s$^{-1}$. Two deficiencies of the model are that it does
not explain the emission one year before peak, and that the
maximum is too broad and bright (\Fig{doublepeak}).

In the third class of models (\Sect{multiple}), SN 1961V was only a
piece of a longer and probably much brighter SN that went largely
unobserved (\Fig{multiple} and \Fig{singleun}) before July 1961. The
narrow width of the peak here is attributed to the collision of
geometrically thin shells. Continuing interaction between shells can
keep the SN bright for years, although ultimately some sort of
interaction with a pre-SN wind would also contribute. This
model, uniquely, allows for the existence of bright emission related
to pulsational activity a year before the main event.

None of these models is perfect. Most of the single-event models are
too bright (e.g., \Fig{singlepeak} and \Fig{doublepeak}). They emit
too much light both at peak and integrated over time. A possible
exception might be Model A123p65 (\Fig{singlepeak}). That model works
if about 40 days of bright emission was missed prior to the
SN's discovery, but SN~1961V was actually unobservable during
that time period, making this a very plausible model (\Sect{absence}).
A still weaker explosion than $4 \times 10^{50}$ erg might also
help. Compare the integrated light output in the observed $\sim$1000
day event (\Tab{61vlite}), which is about $2.4 \times 10^{49}$ erg,
with the higher 1000 day bolometric light output of the models
(\Tab{tmodels}). Part of the discrepancy is that the observations may
have missed some early bright phases of the model's light curve.
Again, though, SN~1961V was unobservable or difficult to observe
before its discovery in July 1961, making this also a plausible model.
Another is that observers at the time generally reported photographic
(blue sensitive) magnitudes, not bolometric magnitudes, and so there
still might be a problem at the factor-of-two level.  Finally, the
models are 1D and nature is not, and so very sharp luminosity spikes
or dips in the models are likely to be somewhat smoother in reality.

The brightest peak in the single event models is also broader than in
the observations. This could reflect too large an envelope mass. A
mass less than the 10 \Msun \ ejected by Model A123p65 would have a
shorter diffusion time and could still satisfy velocity constraints,
but Model A123p7 (not illustrated) had difficulty sustaining steady
emission for 150 days. A perfect single-event fit, or even one better
than the two shown in \Fig{singlepeak}, has been elusive. Perhaps this
speaks in favor of long complex events like \Fig{multiple}, but
implies a very bright SN prior to July, 1961.

What can be done to clarify this situation and discriminate among the
three sets of models?  Most straightforward, though not easy, nuclear
cross section measurements could refine the carbon abundance expected
for such massive stars after helium burning. Here it is a
parameter. In nature, it is not. Better treatments of radiation
transport can clarify the spectrum and bolometric correction for each
successful model. This might not be as easy as it seems, since one
cannot assume a coasting configuration when shock waves are being
repeatedly launched and shells are colliding. Most advanced radiation
transport codes, so far, assume a coasting configuration. Observational
studies of the remnant might discover, or rule out the tens of solar
masses of slow moving oxygen expected in the PISN models or even in
ordinary SNe.  The debate about whether there is still a
star-like object in the site of SN 1961V can continue, but perhaps
with a different interpretation, since a surviving star and a SN are
not incompatible. Searches for high-energy emission from a newly
formed 50 \Msun \ black hole with luminosity near the Eddington limit
might also be carried out. A better understanding of the physics
involved in LBVs is needed in order to address the precursor
variability and mass loss. Rotating PPISN models might be considered
as a way of potentially further weakening the pair instability 
\citep{Mar20,Woo21}.

If SN 1961V was indeed a pulsational SN, there are many interesting
implications for stellar evolution. First, it would confirm that PPISN
do actually happen in nature, and that uncertain mass loss and
reaction rates do not conspire to rule out their existence. Events
like SN 1961V are rare, but they also sample a narrow range of masses,
with helium cores near 50 \Msun. Other varieties of PPISN (and PISN)
should also exist and be sought \citep{Woo17}. The inferred low
metallicity for the SN 1961V site (one-third solar?) suggests a
metallicity dependence in mass-loss rate that may favor the
preservation of very high mass at death, which is consistent with
modern reductions in mass-loss rate prescriptions. For a roughly 110
\Msun \ star with 1/3 \Zsun \ to die with a mass of $\sim$60 \Msun
\ is informative. Finally, this would indicate that a likely route
exists for producing black holes around 40-50 \Msun, many of which
have been found by LIGO.

While the emphasis here is on SN~1961V, we are aware that other
similar events, e.g., SN~2009ip \citep{Mau13,Pas13}, have been
observed and even attributed to PPISN. We save the modeling of these
events for another day, but note in passing that the high velocities
observed in SN 2009ip, both before and during the 2012 outbursts, may
be difficult to accommodate in a PPISN framework without additional
embellishments.  We doubt that all ``supernova impostors'' are PPISNe,
because pair instability supernovae are too infrequent and SN~1961V
was unusually luminous and energetic, but we hope to have made a
compelling case at least one was.

\acknowledgements


\end{document}